\documentclass[twocolumn]{aastex62}

\newcommand{\msun}{M_{\odot}}

\newcommand{\kms}{km~s$^{-1}$}
\newcommand{\mhi}{$M_{\rm HI}$}

\newcommand{\hi}{H{\sc\,i}}

\graphicspath{{./}{figures/}}

\received{January 27, 2020}
\revised{September 7, 2020}
\accepted{December 29, 2020}
\submitjournal{ApJ}

\shorttitle{VLA Imaging of HUDs}
\shortauthors{Gault et al.}

\begin{document}

\title{VLA Imaging of \hi-bearing Ultra-Diffuse Galaxies from the ALFALFA Survey}

\author[0000-0002-2492-7973]{Lexi Gault}
\affiliation{Department of Physics and Astronomy, Valparaiso University,  1610 Campus Drive East, Valparaiso, IN 46383, USA}

\correspondingauthor{Lukas Leisman}
\email{luke.leisman@valpo.edu}

\author[0000-0001-8849-7987]{Lukas Leisman}
\affiliation{Department of Physics and Astronomy, Valparaiso University,  1610 Campus Drive East, Valparaiso, IN 46383, USA}

\author[0000-0002-9798-5111]{Elizabeth A. K. Adams}
\affiliation{ASTRON, Netherlands Institute for Radio Astronomy, Oude Hoogeveensedijk 4, 7991 PD Dwingeloo, The Netherlands}
\affiliation{Kapteyn Astronomical Institute, University of Groningen, Landleven 12, 9747 AD, Groningen, The Netherlands}

\author[0000-0001-5175-939X]{Pavel E. Mancera Pi\~{n}a}
\affiliation{Kapteyn Astronomical Institute, University of Groningen, Landleven 12, 9747 AD, Groningen, The Netherlands}
\affiliation{ASTRON, Netherlands Institute for Radio Astronomy, Oude Hoogeveensedijk 4, 7991 PD Dwingeloo, The Netherlands}

\author[0000-0001-8530-7543]{Kameron Reiter}
\affiliation{Department of Physics and Astronomy, Valparaiso University,  1610 Campus Drive East, Valparaiso, IN 46383, USA}



 

\author[0000-0002-3222-2949]{Nicholas Smith}
\affiliation{Department of Astronomy, Indiana University, 727 East
 Third Street, Bloomington, IN 47405, USA}

\author[0000-0002-3501-8396]{Michael Battipaglia}
\affiliation{Department of Physics and Astronomy, Valparaiso University,  1610 Campus Drive East, Valparaiso, IN 46383, USA}

\author[0000-0002-1821-7019]{John M. Cannon}
\affiliation{Department of Physics \& Astronomy, Macalester College, 1600 Grand Avenue, Saint Paul, MN 55105}

\author[0000-0002-0447-3230]{Filippo Fraternali}
\affiliation{Kapteyn Astronomical Institute, University of Groningen, Landleven 12, 9747 AD, Groningen, The Netherlands}

\author[0000-0001-5334-5166]{Martha P. Haynes}
\affiliation{Cornell Center for Astrophysics and Planetary Science, Space Sciences Building, Cornell University, Ithaca, NY 14853, USA}

\author{Elizabeth McAllan}
\affiliation{Department of Physics and Astronomy, Valparaiso University,  1610 Campus Drive East, Valparaiso, IN 46383, USA}

\author[0000-0002-0786-7307]{Hannah J. Pagel}
\affil{Department of Astronomy, Indiana University, 727 East
 Third Street, Bloomington, IN 47405, USA}

\author[0000-0001-8283-4591]{Katherine L. Rhode}
\affiliation{Department of Astronomy, Indiana University, 727 East
 Third Street, Bloomington, IN 47405, USA}

\author[0000-0001-8483-603X]{John J. Salzer}
\affiliation{Department of Astronomy, Indiana University, 727 East
 Third Street, Bloomington, IN 47405, USA}

\author{Quinton Singer}
\affiliation{Department of Physics \& Astronomy, Macalester College, 1600 Grand Avenue, Saint Paul, MN 55105}



\begin{abstract}

Ultra-diffuse galaxies have generated significant interest due to their large optical extents and low optical surface brightnesses, which challenge galaxy formation models. 
Here we present resolved synthesis observations of 12 \hi-bearing ultra-diffuse galaxies (HUDs) from the Karl G. Jansky Very Large Array (VLA), as well as deep optical imaging from the WIYN 3.5-meter telescope at Kitt Peak National Observatory. We present the data processing and images, including total intensity \hi\ maps and \hi\ velocity fields. The HUDs show ordered gas distributions and evidence of rotation, important prerequisites for the detailed kinematic models in \cite{mancerapina19b}. We compare the \hi\ and stellar alignment and extent, and find the \hi\ extends beyond the already extended stellar component 
and that the \hi\ disk is often misaligned with respect to the stellar one, emphasizing the importance of caution when approaching inclination measurements for these extreme sources.
We explore the \hi\ mass-diameter scaling relation, and find that although the HUDs have diffuse stellar populations, they fall along the relation, with typical global \hi\ surface densities. This resolved sample forms an important basis for more detailed study of the \hi\ distribution in this extreme extragalactic population.

\end{abstract}



\section{Introduction} 
\label{sec:intro}

Very low surface brightness (vLSB) galaxies are essential to our understanding of questions ranging from galaxy formation (e.g., \citealp{agertz16a}) to cosmology (e.g., \citealp{giovanelli16a}), but can be difficult to study at optical wavelengths. 
While vLSB galaxies have been studied for decades (e.g., \citealp{disney76a}; \citealp{sandage84a}; \citealp{ellis84a}; see also reviews by, e.g., \citealp{bothun97a}; \citealp{impey97a}), 
more recent advances in low surface brightness detection techniques (see \citealp{abraham14a,greco18a}) have revealed increased numbers of vLSB galaxies, with particular interest in a subset with stellar masses of dwarf galaxies ($\sim10^8\msun$) but radii comparable to Milky Way sized (L$_{\star}$) galaxies (half light radii of several kpc; e.g., \citealp{vandokkum15a}). Dubbed ``ultra-diffuse" galaxies (UDGs), these galaxies have generated significant attention for, among other things, their potentially extreme dark matter properties (e.g., \citealp{vandokkum16a,vandokkum18a}; see also \citealp{trujillo19a}), and their implications for galaxy formation models, since they make up a non-negligable fraction of the total galaxy population \citep{jones18a, prole19a, danieli19a}.

Galaxies fitting this loose definition of ``ultra-diffuse" have been detected in cluster environments (e.g., \citealp{koda15a, mancerapina19a}), and the field (e.g., \citealp{roman16b, greco18a}), though whether these are completely analogous populations remains unclear.
To date, the distance measurements for most isolated UDGs come from neutral hydrogen (\hi) redshifts, since  optical spectra are difficult to obtain at these low surface brightnesses, and isolated galaxies tend to be gas-rich. 
For example, the Arecibo Legacy Fast ALFA blind \hi\ survey (ALFALFA; \citealp{giovanelli05a}; \citealp{haynes18a}) revealed 253 galaxies \citep{leisman17a, janowiecki19a} out of its over 31,000 extragalactic detections with r$_{r,{\rm eff}}>1.5$~kpc, and $\left< \mu(r,r_{\rm eff}) \right> >24$~mag~arcsec$^{-2}$. 
In addition to having readily obtained redshift measurements, these \hi-bearing ultra diffuse sources (HUDs) provide an opportunity to probe both their gas contents and their dynamics through \hi\ measurements. Indeed, \cite{leisman17a} and \citet{janowiecki19a} note that HUDs appear to have narrow \hi\ velocity widths and elevated gas fractions relative to other \hi\ selected populations of similar \hi\ mass. 

However, these and other single dish \hi\ observations of HUDs (e.g., \citealp{papastergis17a,spekkens18a}) do not resolve these sources, leaving the \hi\ radii, density, and rotation poorly constrained. Though \cite{leisman17a} present resolved synthesis imaging of three HUDs, which tentatively suggested lower than typical gas densities and rotation speeds, their small sample size prevented them from drawing more general conclusions.

Here we present resolved \hi\ and deep optical observations of additional sources from \cite{leisman17a}, that expand the resolved sample, and allow for more robust analysis and broader conclusions. 
We give a detailed presentation of the data and observations, and focus our analysis on the \hi\ mass-diameter relation, and the diffuseness of the gas in relation to the diffuseness of the stellar population.
Analysis of the rotation of the HUDs in a subset of this expanded sample, including their position off the baryonic Tully-Fisher relation \citep{mancerapina19b}, and their angular momentum content \citep{mancerapina20a} are presented elsewhere.   

This paper is organized as follows. In Section \ref{sec:observations} we discuss the optical and \hi\  observations of our sample, and in Section \ref{sec:results} we present the results of those observations. We discuss where the HUDs fall on the \hi\ mass-diameter relation and the implications of such in Section \ref{sec:discussion}, and then present our conclusions in Section \ref{sec:conclusion}. 
For all calculations we assume $H_0 = 70$~\kms~Mpc$^{-1}$,
$\Omega_m=0.3$, and $\Omega_\Lambda=0.7$.

\section{Observations and Analysis}
\label{sec:observations}

\begin{deluxetable*}{cccccccccc}
\tablecaption{Observation Details \label{table:observations}}
\tablecolumns{9}
\tablewidth{0pt}
\tablehead{
\colhead{AGC ID\tablenotemark{a}} &
\colhead{OC RA\tablenotemark{b}} &
\colhead{OC Dec\tablenotemark{b}} &
\colhead{WIYN\tablenotemark{c}} &
\colhead{VLA\tablenotemark{d}} &
\colhead{VLA\tablenotemark{e}} &
\colhead{$t_{\rm HI}$\tablenotemark{f}} &
\colhead{$F_{\text{VLA}}/$\tablenotemark{g}} &
\colhead{$\sigma$ cube\tablenotemark{h}} &
\colhead{Beam\tablenotemark{i}}\\[-0.8em]
\colhead{} &
\colhead{J2000} &
\colhead{J2000} &
\colhead{Date} &
\colhead{Date} &
\colhead{Config} &
\colhead{hours} &
\colhead{$F_{\text{ALFA}}$} &
\colhead{mJy~bm$^{-1}$} &
\colhead{\arcsec~$\times$~\arcsec}
}
\startdata
114905 & 21.3271	& 7.3603 & Oct 2016  & Jul 2017  & C & 6  & 0.98 &0.82 & 14.5$\times$13.0\\
122966 & 32.3708 & 31.8528 & Nov 2013 (pODI)   &Aug 2017  & C & 6  & 1.07 &0.88 & 16.7$\times$13.6\\
198596 &  147.0300 & 16.2606 &Apr 2018     & Aug 2017    & C & 4 & --   & -- & -- \\
219533 & 174.9867 & 16.7214& Mar 2017, Apr 2018  & Dec 2014& C & 6 &  1.05 &0.79 & 14.9$\times$13.6\\
229110 & 191.5362 & 28.7508 & Apr 2018   & Aug 2017        & C & 6 & 0.87 &0.77 & 15.8$\times$13.8\\
238764 &204.9063 & 6.9961& Apr 2018    & Aug 2017       & C & 6 &  0.55 &0.83 & 18.0$\times$14.4\\
248937 & 216.4779 & 12.9189& Apr 2018    & Aug 2017      & C & 6 & 0.98 &1.12 & 15.9$\times$13.9 \\
248945 &221.7479	& 13.1697 & Apr 2018    & Aug 2017      & C & 6&  0.59 & 0.85 & 18.1$\times$14.1 \\
334315 &350.0492 & 22.4019& Sept 2017  & Jun 2017   & C & 6 &  1.17 &0.76 & 15.8$\times$13.9\\
748738 &346.2167 & 14.0181& Oct 2016, Sept 2018  & Jul 2017& C & 6 &  1.01 & 0.79 & 17.1$\times$14.5 \\
749251 &113.7513 & 26.6589& Apr 2018   & Apr 2017, Aug 2017       & C, D & 6, 2&  0.94 & 0.73 & 16.4$\times$15.2 \\ 
749290 & 139.0046 & 26.6497 & Apr 2018    & 
\begin{tabular}{@{}c@{}}Mar 2017\\ Apr 2017, Jul 2017\end{tabular} & C, D & 6, 2&  1.01 & 0.83 & 17.6$\times$14.5
\enddata
\tablenotetext{a}{Galaxy identifier from the Arecibo General Catalog}
\tablenotetext{b}{Galaxy coordinates based on the centroid of the optical component of the emission, in decimal degrees.} 
\tablenotetext{c}{Observation dates for optical observations with the WIYN 3.5m at KPNO.}
\tablenotetext{d}{Observation dates for radio observations with the VLA.}
\tablenotetext{e}{VLA configuration(s) used for the \hi\ imaging.}
\tablenotetext{f}{The number of hours observed with the VLA in the given configurations, including time on calibrators.}
\tablenotetext{g}{Ratio of the recovered VLA flux to the measured ALFALFA flux.}
\tablenotetext{h}{RMS noise in the final cleaned VLA images.}
\tablenotetext{i}{Major and minor axes of the \hi\ beam in the final cleaned VLA images.}
\end{deluxetable*}

\subsection{HI data}
\label{sec:data.hi}

We observed 11 sources using the  Karl G. Jansky Very Large Array (VLA) in March-August of 2017 (proposal 17A-210; P.I. Leisman), and include one additional source (AGC~219533) observed during a previous set of VLA observations (proposal 14B-243;  
\citealp{leisman17a}), for a total sample of 12 galaxies (Table \ref{table:observations}). Two of the sources (AGC~334315 and AGC~122966) also have data from the Westerbork Synthesis Radio Telescope (WSRT) presented in \cite{leisman17a}. We note that these sources were chosen from early versions of the \cite{leisman17a} catalog, which required r$_{r,{\rm eff}}>2$~kpc, and $\left< \mu(r,r_{\rm eff}) \right> >24$~mag~arcsec$^{-2}$, as measured SDSS data, nearest neighbor separations $>350$~kpc on the sky and $>$500~\kms, and distances $<$120~Mpc. The observed sample size was further restricted by requiring that sources had \mhi$>10^8\msun$, and that they fell in fields where mid-depth GALEX data was available, and did not have missing or bad SDSS data (due to, e.g., nearby bright stars). The GALEX and SDSS data requirements are approximately random across the region covered by ALFALFA, thus the synthesis sample should be approximately representative of the broader ALFALFA sample from \cite{leisman17a}.

We observed all sources in ``C" configuration, and two sources in ``D" configuration due to scheduling pressure constraints; details for each galaxy are listed in Table \ref{table:observations}. We typically observed the sources for three two-hour observing blocks in C-configuration, or for two one-hour observing blocks in D-configuration. We observed the nearest of the standard flux calibrators, 3C48, 3C147, or 3C286, for approximately 10 minutes at the beginning or end of each observation and observed the nearest appropriate phase calibrator from the VLA catalog with a conservative cadence of approximately 2 minutes of phase calibration for every 18 minutes of time on source.
We used the WIDAR correlator in dual polarization mode with a single 8~MHz wide sub-band with 1024 channels, giving a native channel width of 1.7~\kms, which we smooth to a velocity resolution of 4~\kms. 

The data reduction process followed standard methods in the Common Astronomy Software Application (CASA; \citealp{mcmullin07a}). Flagging of the visibilities was done by hand, and standard cross-calibration was performed with the primary calibrator used to determine the flux scale and bandpass. The phase calibrator was used to determine the complex gains over the course of the observation, 
and continuum subtraction in the uv plane was applied to each source. 
We created data cubes by combining all calibrated data sets using the CASA task CLEAN in interactive mode. We used
a Briggs robust weighting of 0.5, a cleaning threshold of 3.0 mJy or 3$\sigma$, and the multiscale clean option, creating images with 3\arcsec\ pixels and 4~\kms\ channels. The resulting noise and beam size from each cube is listed in Table 
\ref{table:observations}. 

We note that four of these resulting VLA image cubes  (AGC~114905,AGC~219533, AGC~248945, and AGC~749290) were used in the analysis of \cite{mancerapina19b} (which also include WSRT observations of AGC~334315 and AGC~122966 from \cite{leisman17a}), and six in \cite{mancerapina20a} (the four above, plus AGC~334315 and AGC~122966).

We created moment 0 and moment 1 maps using masked data cubes, with masks calculated at 2$\sigma$ (for moment 0) and 3$\sigma$ (for moment 1) on data cubes smoothed to twice the beam size (we use a higher masking threshold for moment 1 analysis to reduce noise near the edge of the maps), and using only channels in the velocity range of the source. We converted the moment 0 total flux maps to \hi\ column density maps assuming optically thin \hi\ gas that fills the beam, and perform a final masking on the resulting maps (using a mask calculated at the 3$\sigma$ level from a smoothed moment 0 map) to create the images presented below. 
We note that our masking technique is different from that used in \cite{mancerapina19b}, resulting in some visual differences in the resulting moment maps, but having little impact in the measured parameters, as discussed in section \ref{sec:discussion.inclin}.
We extracted spectra from the unmasked data cubes within the spatial region defined by the moment 0 maps, and fit the resulting line profiles to obtain measurements of total flux, redshift, and line width, which we report in Table \ref{table:HIproperties}. We compared the resulting recovered VLA flux to that of ALFALFA as additional verification of our reduction; this ratio is also reported in Table \ref{table:observations}.

We computed \hi\ masses from the VLA flux measurements assuming that the \hi\ is optically thin using the standard formula (e.g., \citealp{roberts75a}): \begin{equation}
M_{\rm \sc{HI}} = 2.36\times 10^5 D^2 \int SdV
\end{equation} where $D$ is the distance in Mpc and $\int SdV$ is the integrated \hi\ line flux in Jy~km~s$^{-1}$. We assumed distances from the ALFALFA catalog \citep{haynes18a} as reported in Table \ref{table:HIproperties}. We calculated uncertainties on the \hi\ mass following \cite{haynes18a} by combining the uncertainty in the distance, the integrated line flux, and a 10\% systematic flux calibration uncertainty. 

We note that we were unable to recover an image of AGC~198596, and thus exclude it from our analysis. One of the three C configuration observations of this source was corrupted and unusable, and the other two observations had extreme RFI, such that over 60\% of the data had to be flagged in both observations. 


\subsection{Optical Data}
\label{sec:data.optical}

We observed 11 of the 12 sources in our sample between 2016 and 2018 using the One Degree Imager (ODI; \citealp{harbeck14a}) on the 3.5-meter WIYN telescope at Kitt Peak National Observatory, and one source in 2013 using ODI when it was partially-populated (pODI; see Table \ref{table:observations}). Each source was observed in the \textit{g} and \textit{r} bands for a total exposure time of 45 minutes per filter. The observations typically consisted of a nine-point dither pattern of 300 second exposures in order to fill in gaps between the CCD detectors.  

We reduced the images using the Quick Reduce \citep{kotulla14a} data reduction pipeline through the One Degree Imager Pipeline, Portal, and Archive (ODI-PPA; \citealt{gopu14a}) science gateway. 
This pipeline performs the following tasks: masks saturated pixels; corrects crosstalk and persistence; subtracts the overscan signal; corrects for nonlinearity; applies the bias, dark, and flat-field corrections; corrects for pupil ghosts; and removes cosmic rays.  We also perform an illumination correction, subtract the sky background, and scale the images to a common flux level. The scaling factors for the images are calculated based on measurements of the peak fluxes in a few dozen bright, unsaturated stars distributed across all areas of the field.  We then stack the scaled images and restore the appropriate background level to the final, stacked image.  The same set of bright stars is again used to measure the typical full width at half-maximum of the point spread function (FWHMPSF) in the images.   The average FWHMPSF for all images in our sample is 0.8\arcsec\ in \textit{g} and 0.9\arcsec\ in \textit{r}, with a range of 0.6\arcsec\ - 1.7\arcsec\ in both bands.

\begin{figure*}[t!]
\includegraphics[width=0.325\textwidth]{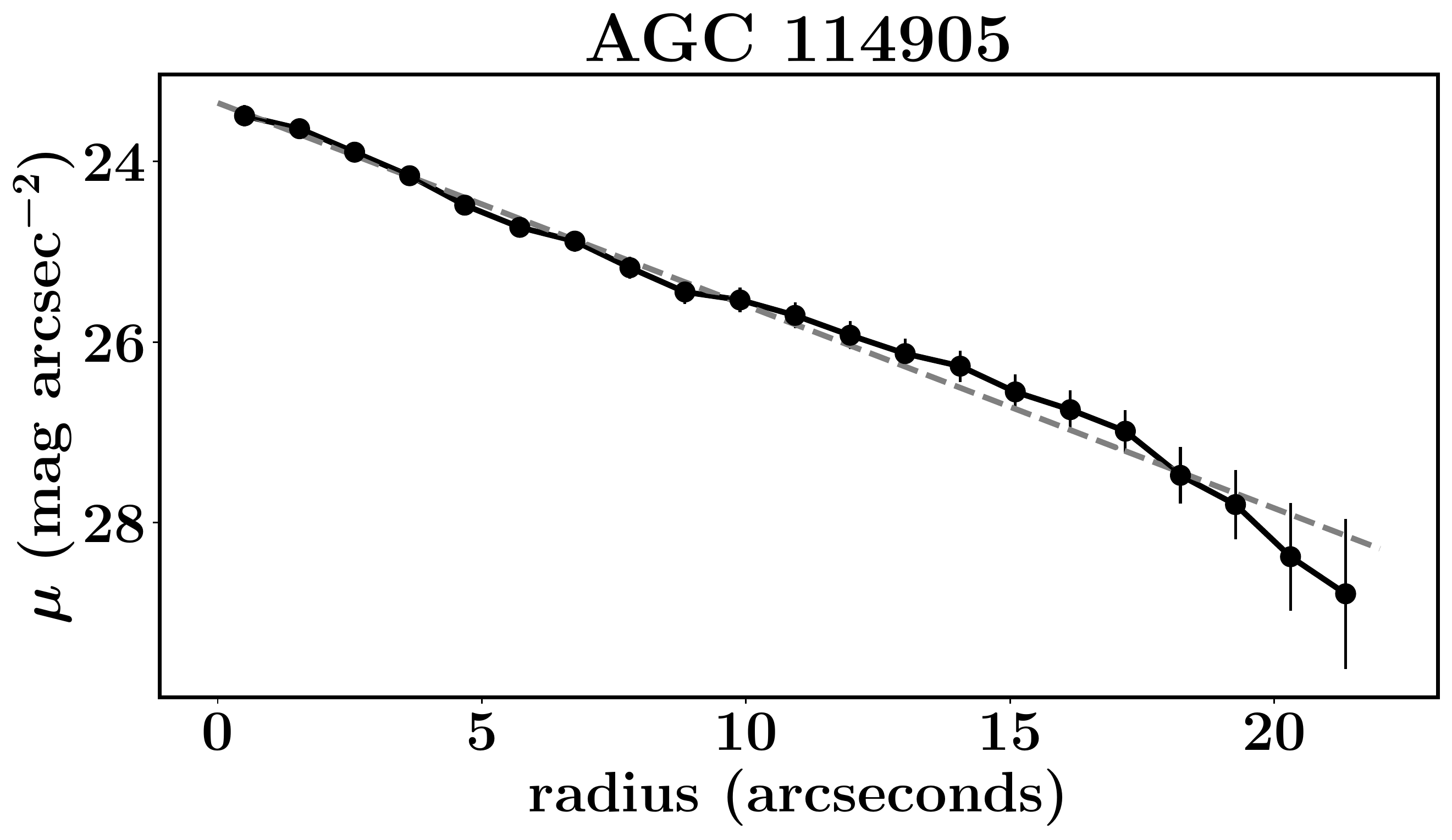}
\includegraphics[width=0.325\textwidth]{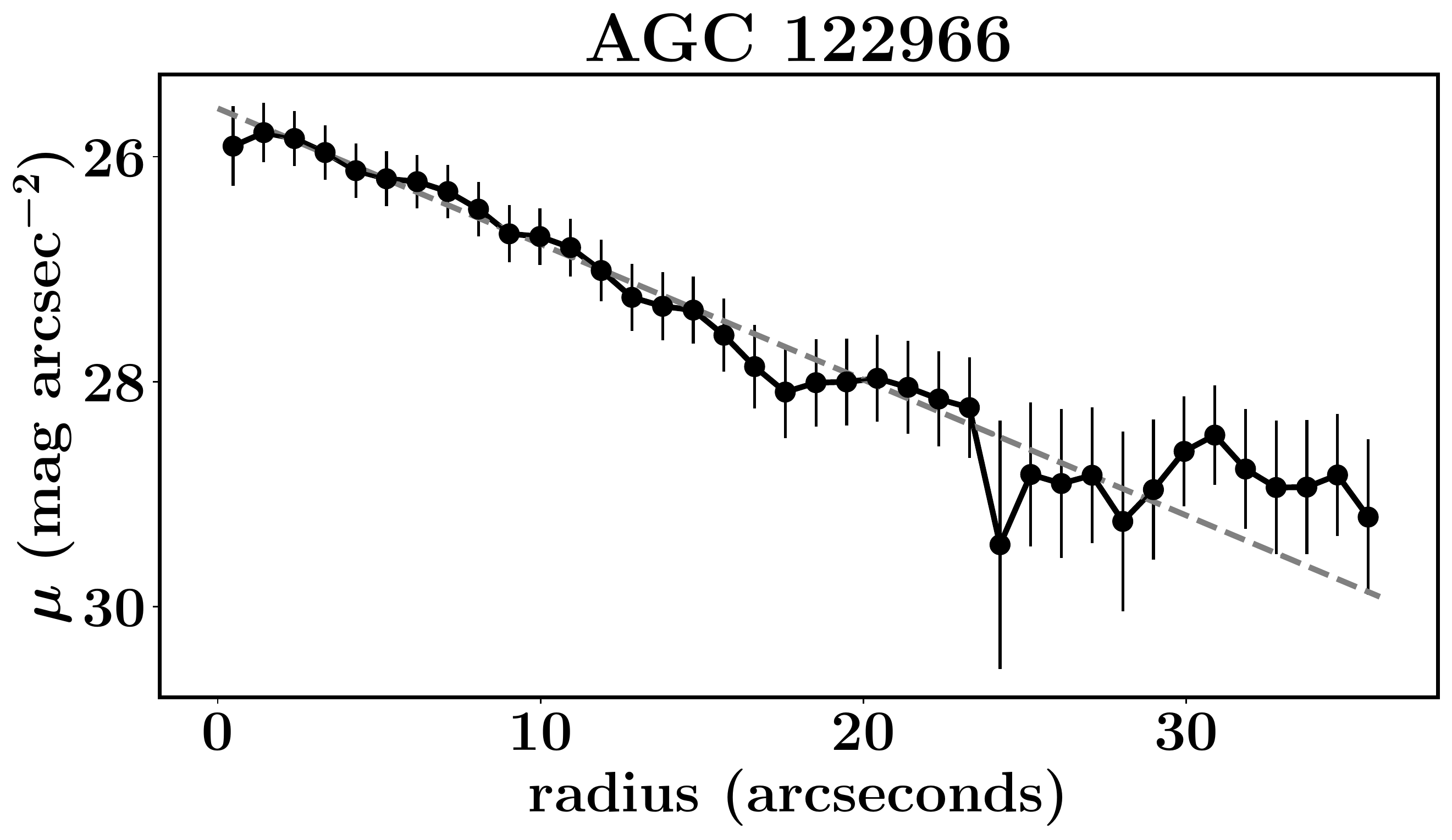}
\includegraphics[width=0.325\textwidth]{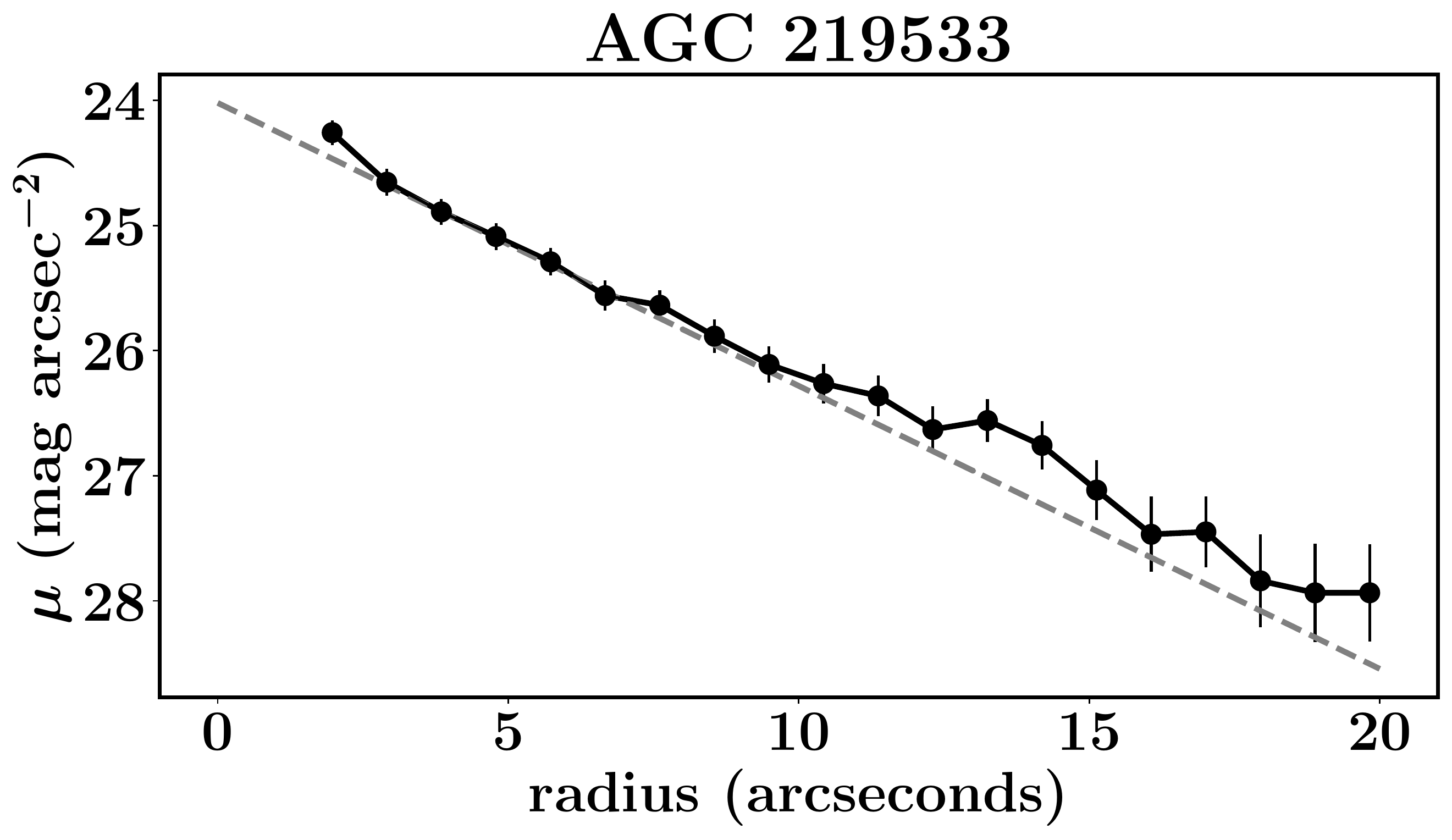}
\includegraphics[width=0.325\textwidth]{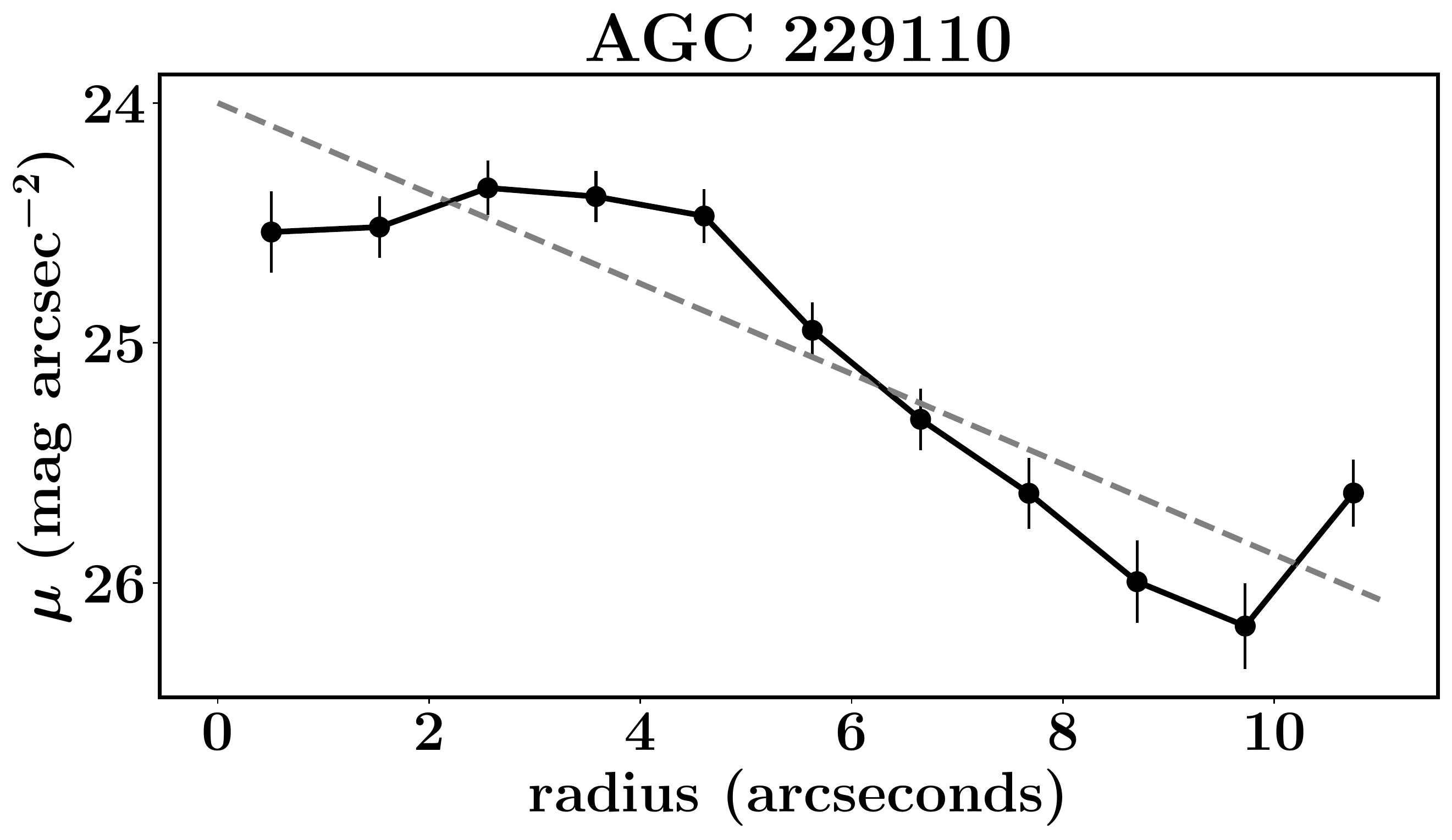}
\includegraphics[width=0.325\textwidth]{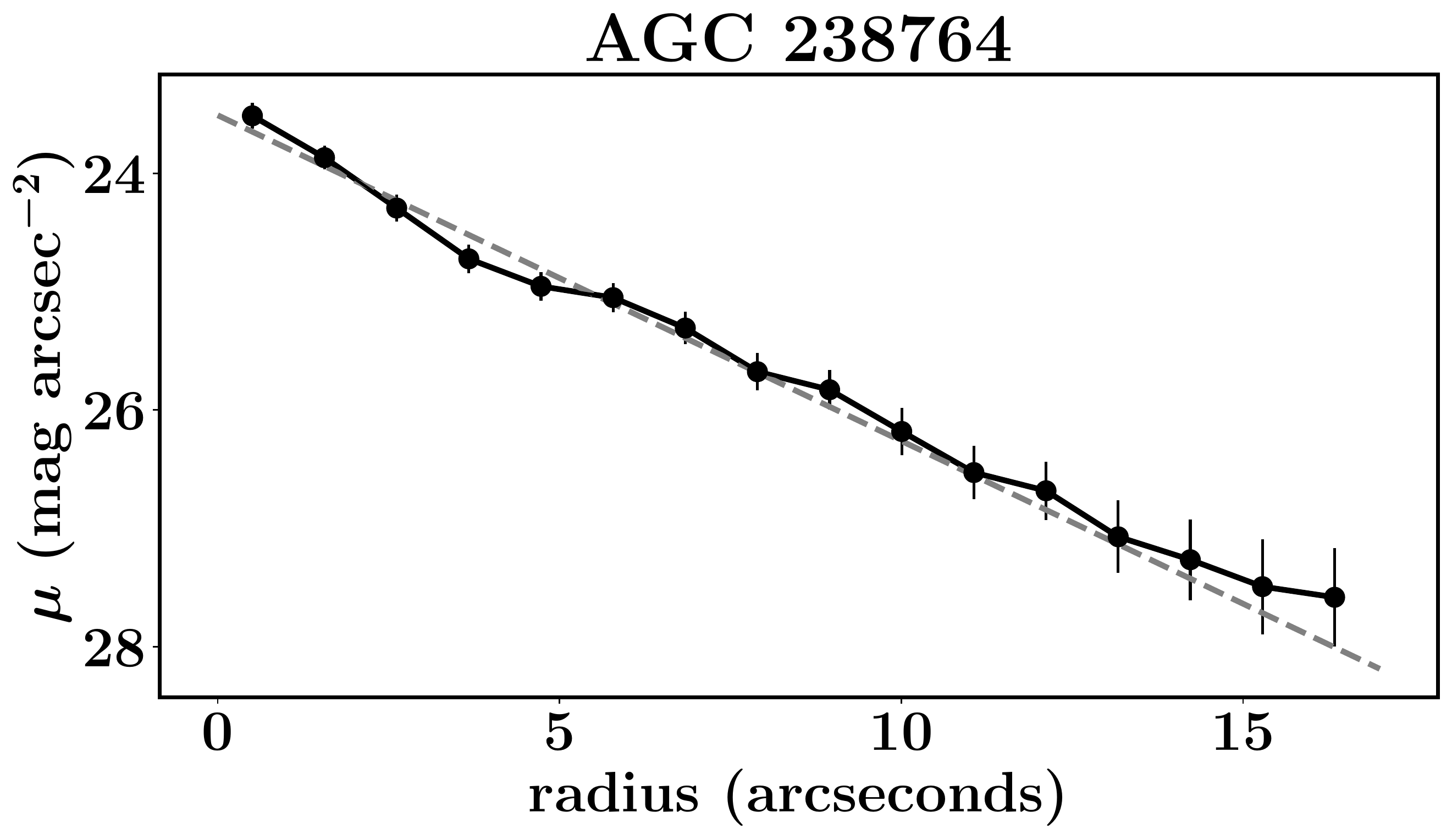}
\includegraphics[width=0.325\textwidth]{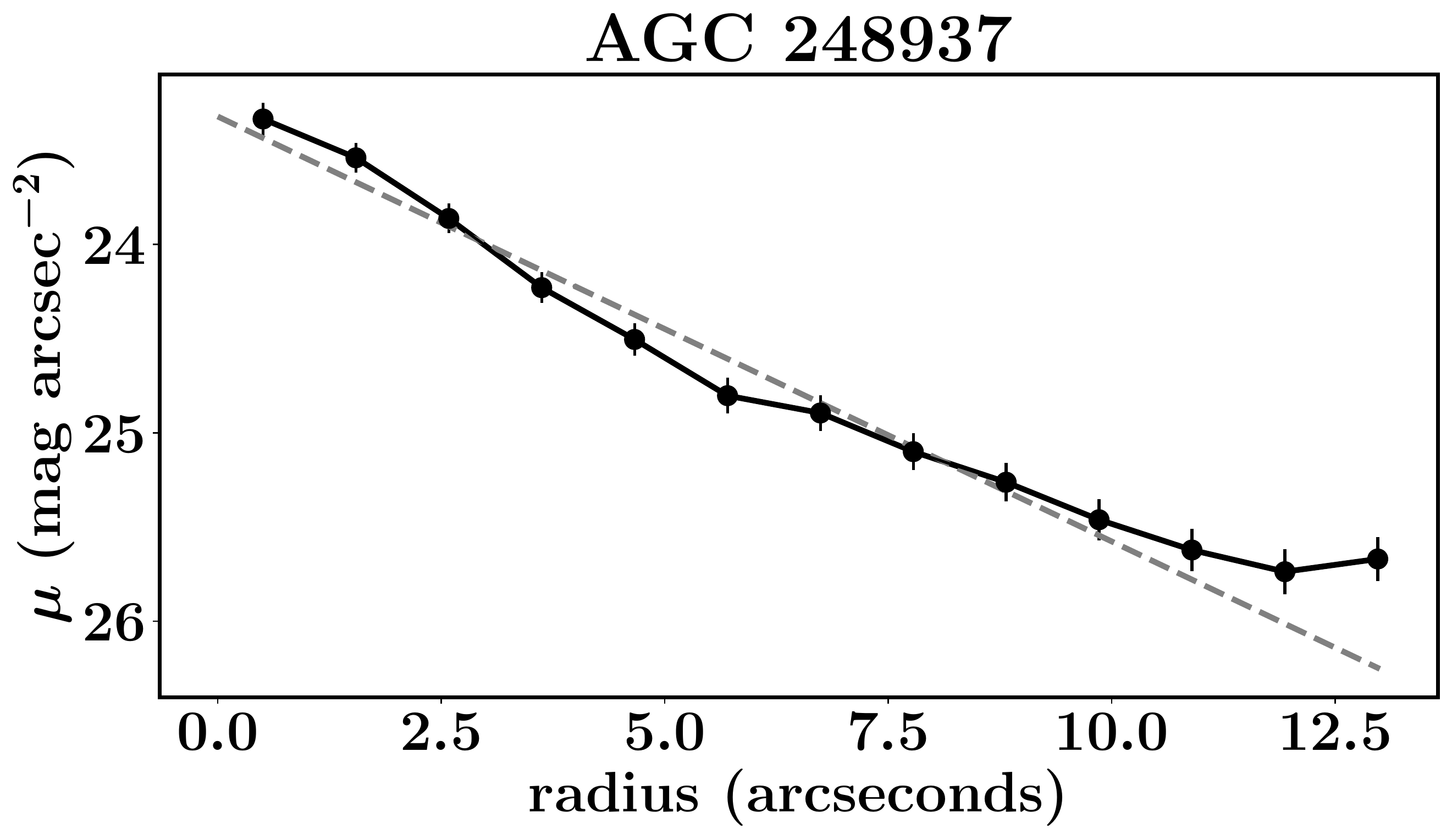}
\includegraphics[width=0.325\textwidth]{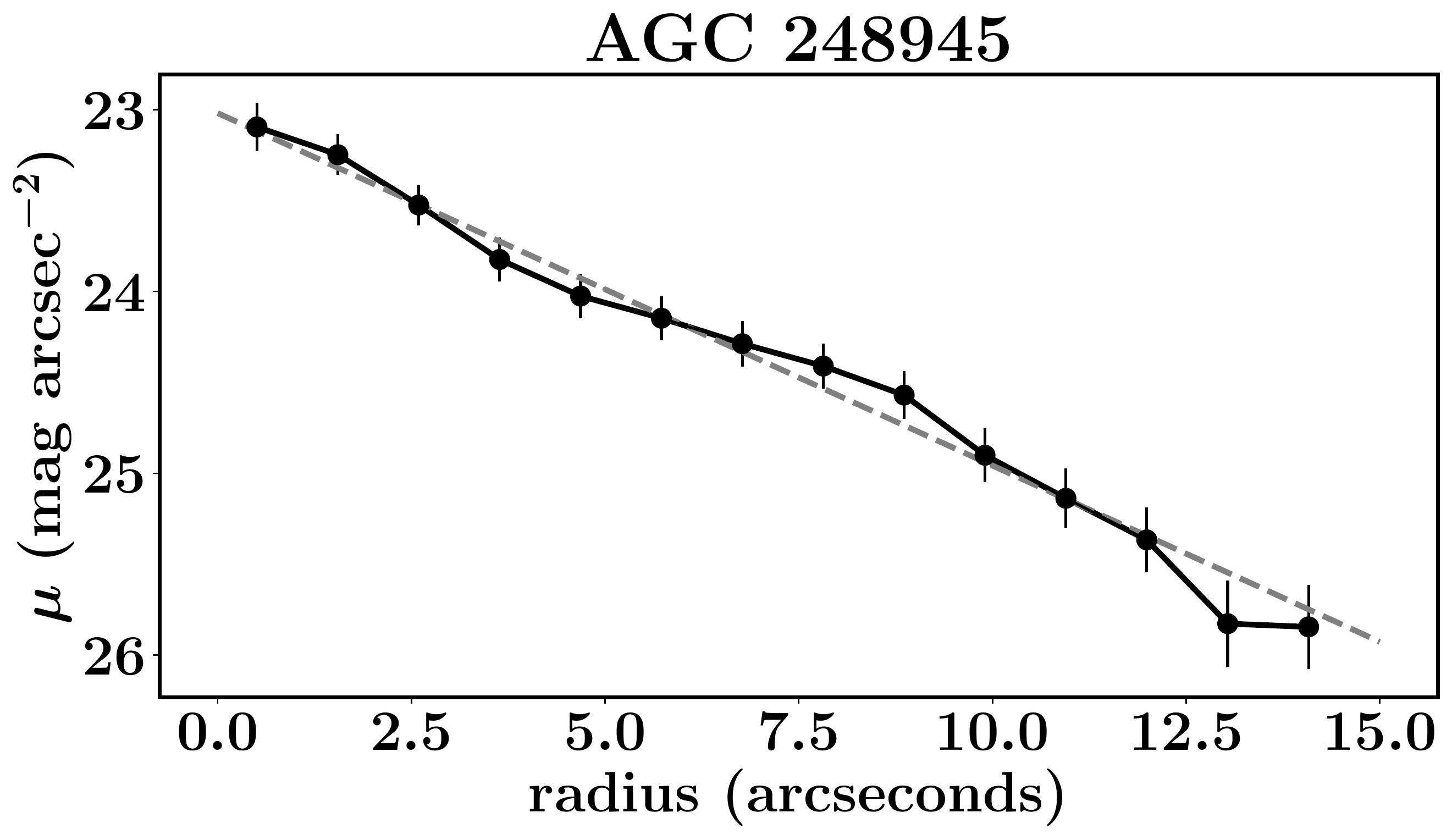}
\includegraphics[width=0.325\textwidth]{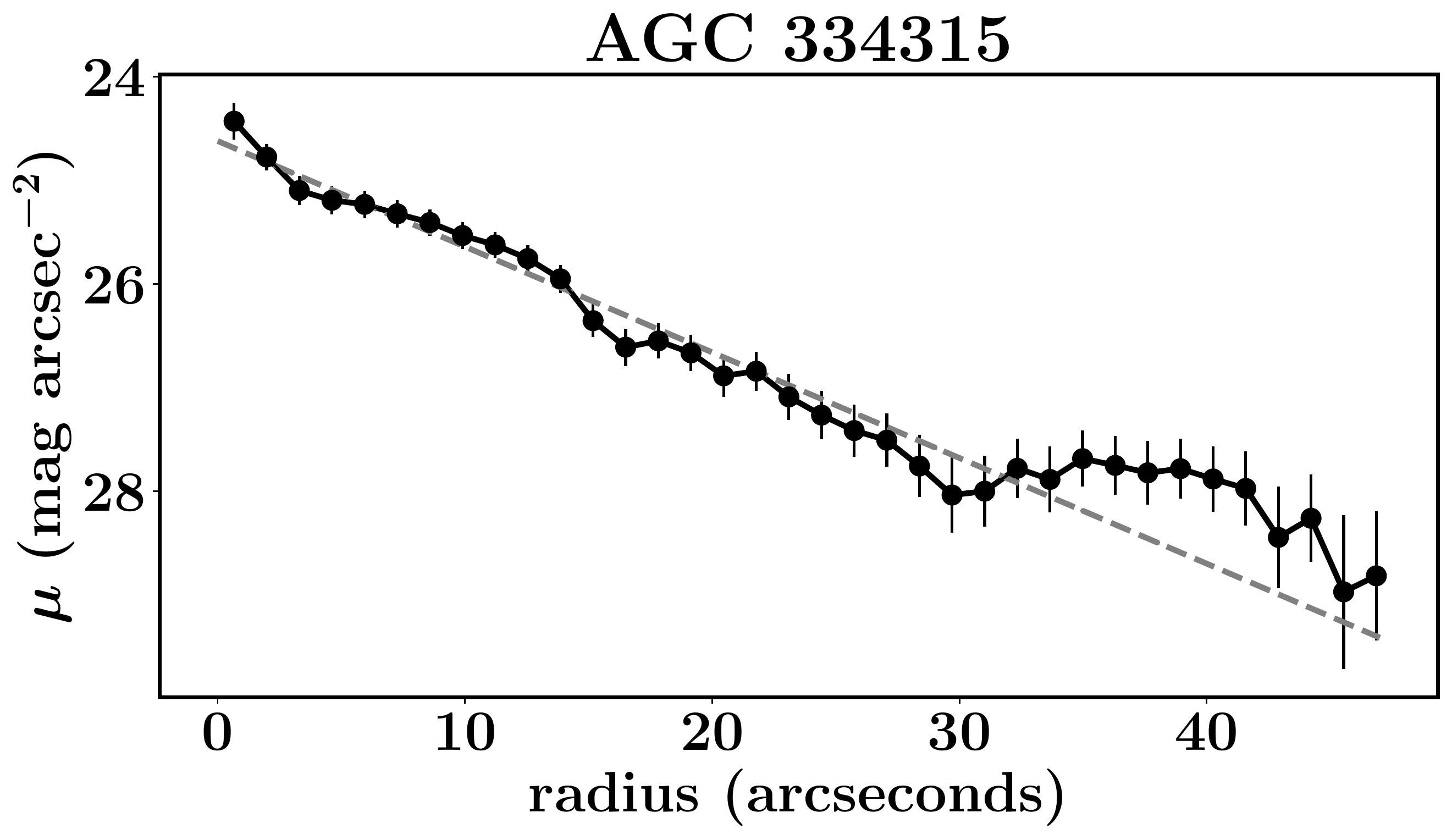}
\includegraphics[width=0.325\textwidth]{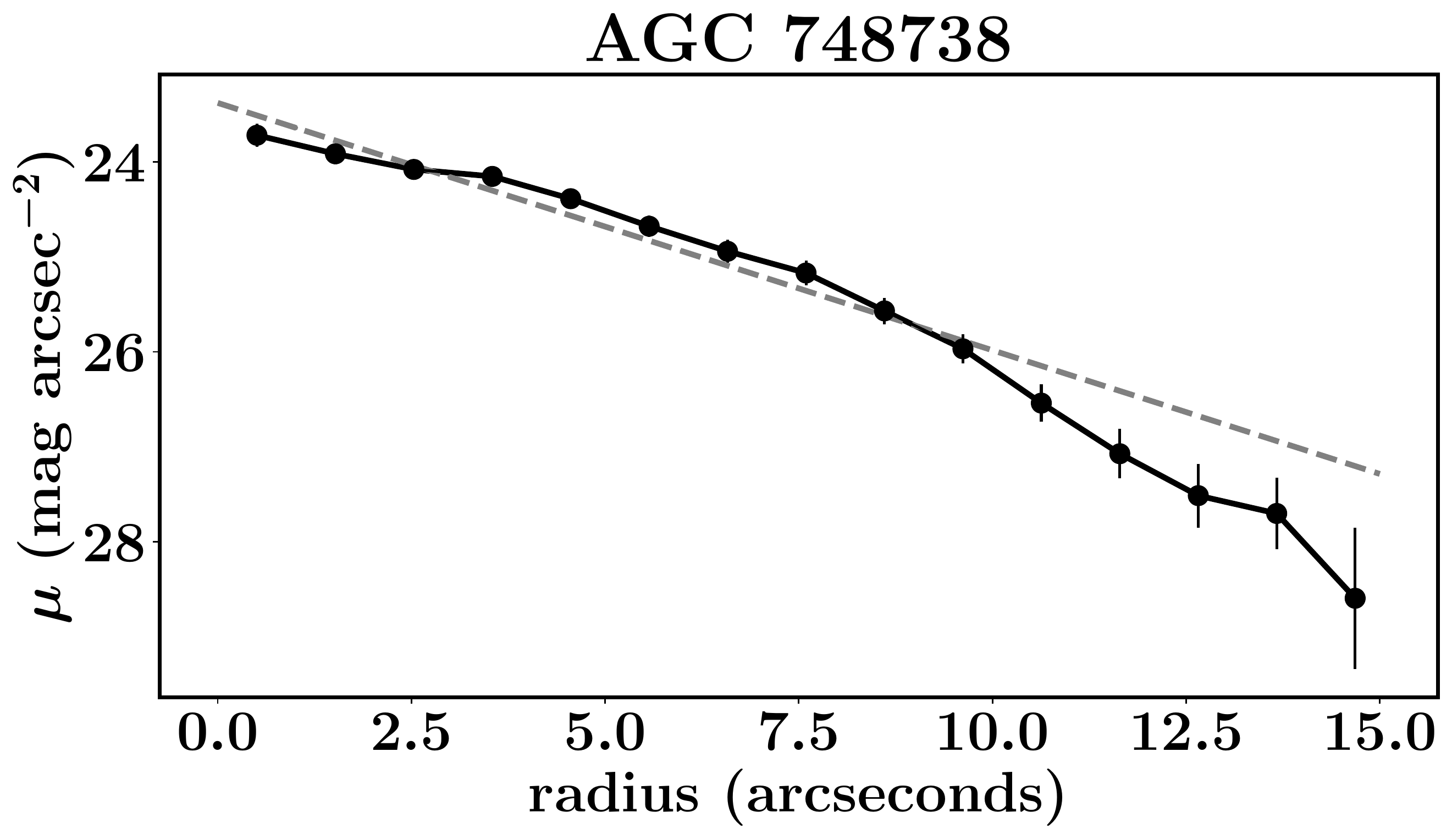}
\includegraphics[width=0.325\textwidth]{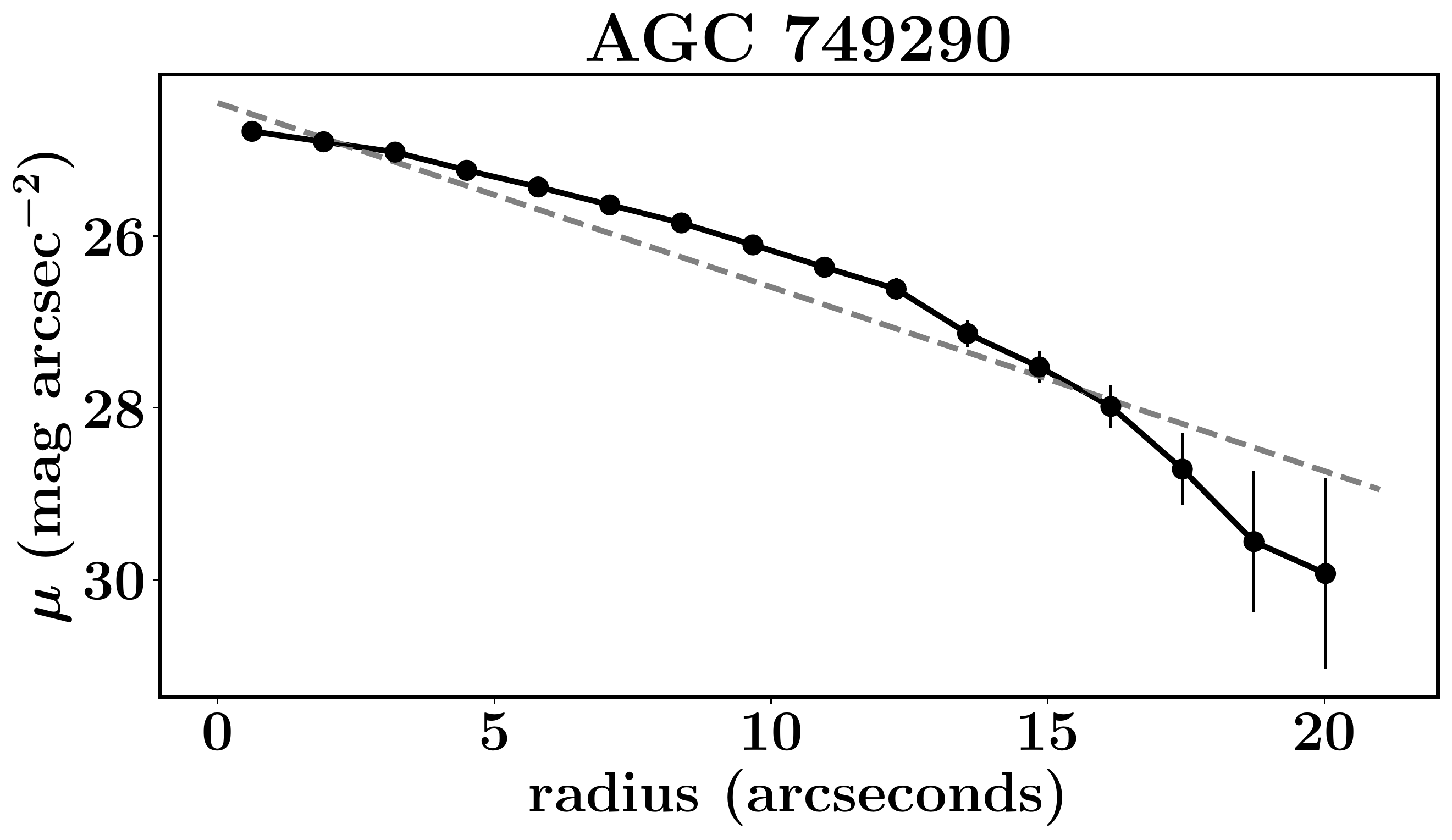}
\includegraphics[width=0.325\textwidth]{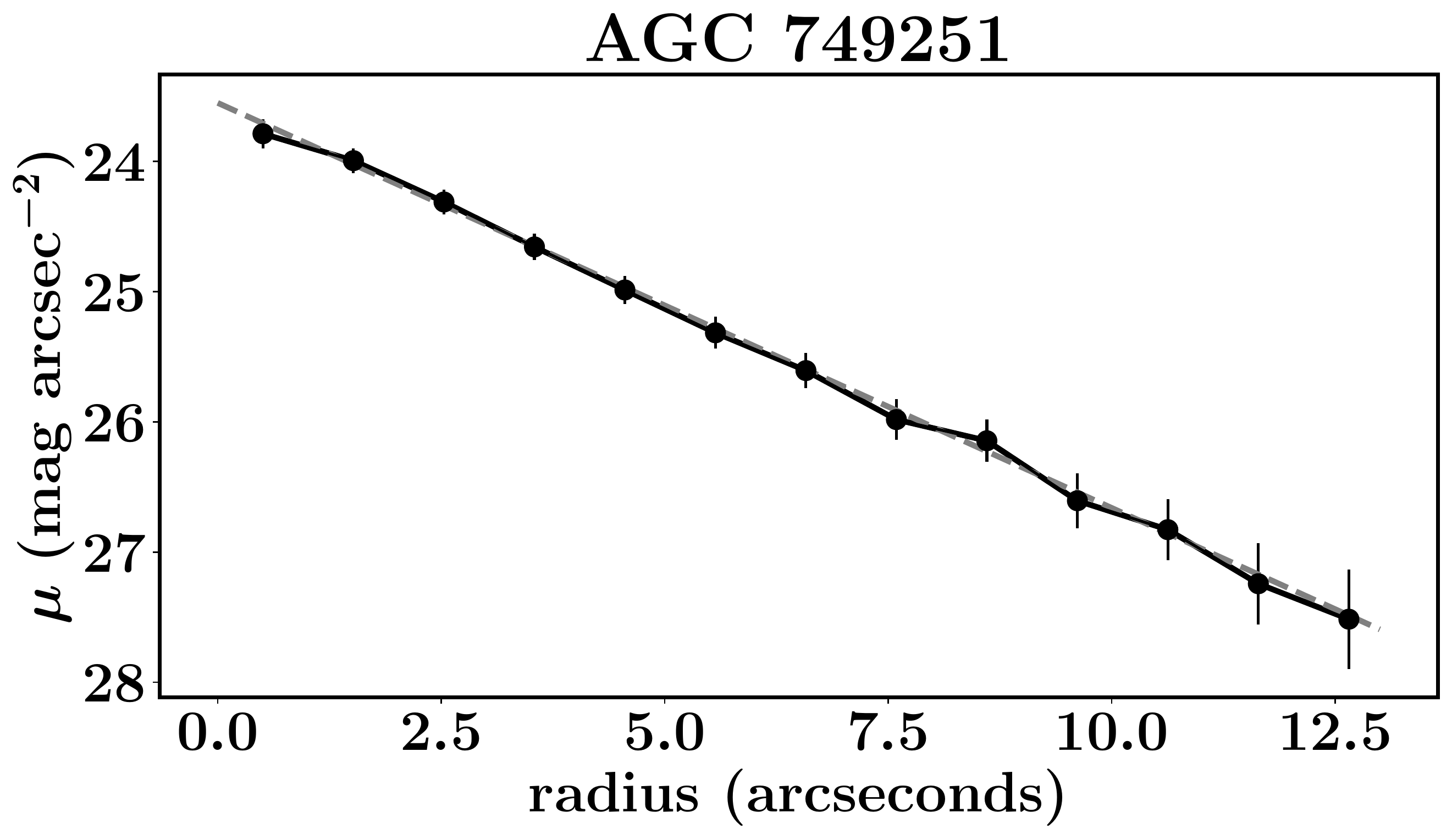}
\vspace{-0.1cm}
\caption{Surface brightness profiles from ellipses fit to $r$ band ODI and pODI images. Data are shown in black, exponential fits are shown with grey dashed lines.
\label{fig:sbprofs}
}
\end{figure*}

We measure simplistic optical surface brightness profiles (Figure \ref{fig:sbprofs}) following the procedure presented in \cite{mancerapina19b}. In brief, we perform background subtraction, noise estimates, and
masking following the procedure detailed in \cite{marasco19a}, and determine position angles and axial ratios using SExtractor. We then fit ellipses using a fixed position angle, centered on optical centers (determined either from SExtractor or the SDSS optical coordinates from the ALFALFA catalog \citep{haynes11a}) of the galaxy. We note that the centroid position from SExtractor and from \citealp{haynes11a} mostly agree, except in cases where the center of light is not the brightest point in the galaxy. The choice of centroid makes very little difference to our measured magnitudes ($\sim$0.02 mag), but can have an impact of $\sim$0.1 mag~arcsec$^{-2}$ on the measured central surface brightness, and of $\sim$0.4~kpc on the effective radii. 

We mask the images in a semi-automatic fashion, modifying the signal-to-noise threshold for applying masks to nearby sources as appropriate. When in doubt, we chose to be conservative with our masking, not masking features that might be part of the galaxy. We then compute total magnitudes, surface brightnesses, and radii using exponential fits to the resulting profiles; these values are given in Table \ref{table:optproperties}. 

We estimate stellar masses using the color-M/L relation from \cite{herrmann16a}. This relation is calibrated for dIrr-like galaxies, which have many similarities in their stellar properties with field UDGs (though see \cite{du20a}, published after original submission of this work, for a discussion of potential systematic offsets). The uncertainties in our colors, magnitudes, and the color-M/L relation result in Gaussian distributions for each stellar mass estimate; the value reported in Table \ref{table:optproperties} corresponds to the median of this distribution, and its uncertainties represent the difference of this median with the 16$^{\rm th}$ and 84$^{\rm th}$ percentiles.

%

%
%
\begin{deluxetable*}{cccccccccc}
\tablecaption{Observed Optical Properties of HUDs \label{table:optproperties}}
\tablecolumns{9}
\tablewidth{0pt}
\tablehead{
\colhead{AGC ID} &
\colhead{$\mu_{g,0}$\tablenotemark{a}} &
\colhead{$r_{\rm eff}$\tablenotemark{b}} &
\colhead{$r_{25}$\tablenotemark{c}} &
\colhead{$g - r$\tablenotemark{d}} &
\colhead{$m_{r}$\tablenotemark{e}} &
\colhead{$M_{r}$\tablenotemark{f}} &
\colhead{$\log(M_{*}/\msun)$\tablenotemark{g}} &
\colhead{A$_{g}$\tablenotemark{h}} &
\colhead{A$_{r}$\tablenotemark{i}}\\[-0.8em]
\colhead{} &
\colhead{mag~arcsec$^{-2}$} &
\colhead{kpc} &
\colhead{kpc} &
\colhead{mag} &
\colhead{mag} &
\colhead{mag} &
\colhead{} &
\colhead{mag}&
\colhead{mag}
}
\startdata
114905 & 23.74$\pm$0.13 &2.99$\pm$0.08& 2.69 & 0.30$\pm$0.12 &18.02$\pm$0.09 & -16.47 $\pm$ 0.17 & 8.30$\pm$0.17 & 0.119 & 0.082\\
122966 &  25.65$\pm$0.23 &6.59$\pm$0.20& 3.27 & -0.10$\pm$0.22 &18.47$\pm$0.15 & -16.49 $\pm$ 0.19 & 7.73$\pm$0.12 & 0.273 & 0.189\\
219533 & 24.14$\pm$0.33 &3.75$\pm$0.13& 2.05 & 0.12$\pm$0.12 &18.62$\pm$0.08 & -16.34 $\pm$ 0.23 & 8.04$\pm$0.12 & 0.077 & 0.053\\
229110 & 24.33$\pm$1.70 &5.26$\pm$0.84& 2.72 & 0.19$\pm$0.11 & 18.80$\pm$0.08 & -16.47 $\pm$ 0.13 & 8.20$\pm$0.13 & 0.037 & 0.026\\
238764 & 23.76$\pm$0.18 &3.34$\pm$0.13& 2.70 & 0.13$\pm$0.11 &18.87$\pm$0.08 &-16.27 $\pm$ 0.13 & 8.06$\pm$0.12 & 0.074 & 0.051\\
248937 & 23.65$\pm$0.29 & 4.62$\pm$0.28 & 4.40 & 0.23$\pm$0.12 & 18.19$\pm$0.07 & -17.24 $\pm$ 0.12 & 8.54$\pm$0.14 & 0.102 & 0.070\\
248945 & 23.38$\pm$0.35 &3.83$\pm$0.15& 4.02 & 0.32$\pm$0.11 &17.35$\pm$0.07 & -17.32 $\pm$ 0.15 & 8.52$\pm$0.17 & 0.063 & 0.044\\
334315 & 24.74$\pm$0.13 &6.32$\pm$0.22& 0.63 & -0.08$\pm$0.18 &18.23$\pm$0.15 &-16.24 $\pm$ 0.21 & 7.93$\pm$0.12 & 0.223 & 0.154\\
748738 & 23.79$\pm$0.41 &1.90$\pm$0.13& 1.56 & 0.16$\pm$0.12 &18.51$\pm$0.09  & -15.77 $\pm$ 0.21 & 7.89$\pm$0.14 & 0.785 & 0.543\\
749251 & 23.99$\pm$0.18 &3.01$\pm$0.05& 2.32 & 0.15$\pm$0.12 &19.02$\pm$0.08  & -16.20 $\pm$ 0.13 & 8.06$\pm$0.12 & 0.138 & 0.095\\
749290 & 24.75$\pm$0.30 &4.00$\pm$0.24& 1.77& 0.17$\pm$0.12 &18.18$\pm$0.08 & -16.83 $\pm$ 0.14 & 8.32$\pm$0.13 & 0.108 & 0.075
\enddata
\tablenotetext{a}{Central surface brightness in the $g$ band, uncorrected for Galactic extinction or surface brightness dimming. Obtained by fitting an exponential profile to the observed $g$ band surface density profile.}
\tablenotetext{b}{Effective radius, assuming an exponential profile, derived from the $r$ band surface brightness profile.}
\tablenotetext{c}{Isophotal radius as measured at the 25~mag~arcsec$^{-2}$ isophote.}
\tablenotetext{d}{Color, corrected for Galactic extinction.}
\tablenotetext{e}{Apparent $r$ band magnitude, uncorrected for Galactic extinction or surface brightness dimming.}
\tablenotetext{f}{Absolute $r$ band magnitude, corrected for Galactic extinction, and including a distance error of 5~Mpc.}
\tablenotetext{g}{Stellar mass, estimated from our (Galactic-extinction corrected) colors and absolute magnitudes, by means of the color - M/L relations from \cite{herrmann16a}.}
\tablenotetext{h,i}{\hspace{6pt} Galactic extinction correction \citep{schlafly11a} from the NASA Extragalactic Database in the $g$ and $r$ bands, respectively.}
\end{deluxetable*}

\begin{deluxetable*}{cccccccccc}

\tablecaption{Observed \hi\ Properties of HUDs \label{table:HIproperties}}
\tablecolumns{10}
\tablewidth{0pt}
\tablehead{
\colhead{AGC ID} &
\colhead{Distance\tablenotemark{a}} &
\colhead{$V_{21}$\tablenotemark{b}} &
\colhead{$W_{50}$\tablenotemark{c}} &
\colhead{$F_{VLA}$} &
\colhead{log($M_{\rm HI}/\msun$)} &
\colhead{$a \times b$\tablenotemark{d}} &
\colhead{$a \times b$\tablenotemark{e}}& 
\colhead{$\sigma_{a}$\tablenotemark{f}}&
\colhead{$N_{\rm peak}$\tablenotemark{g}}\\[-0.8em]
\colhead{} &
\colhead{Mpc} &
\colhead{km~$s^{-1}$} &
\colhead{km~$s^{-1}$} &
\colhead{Jy~km~$s^{-1}$} &
\colhead{} &
\colhead{arcsec $\times$ arcsec} &
\colhead{kpc $\times$ kpc}& 
\colhead{kpc}&
\colhead{$10^{20}~{\rm cm}^{-2}$}
}
\startdata
114905 & 76 & 5429.4$\pm$0.64& 34.1$\pm$1.6 & 0.94$\pm$0.04 & 9.10$\pm$0.05 & 29.3$\times$26.6 & 10.3$\times$9.3 & 1.1 & 5.9 \\
122966 &90 & 6517.4$\pm$0.90& 42.5$\pm$2.2 &0.57$\pm$0.02& 9.03$\pm$0.05 & 24.5$\times$22.1 & 9.9$\times$7.8 & 1.3 & 5.6 \\
219533 & 96& 6381.5$\pm$1.8& 60.7$\pm$5.0 & 0.73$\pm$0.04 & 9.24$\pm$0.06 & 29.1$\times$22.3 & 12.8$\times$9.6 & 1.4 & 7.5 \\
229110 & 112& 7552.8$\pm$1.6& 45.3$\pm$4.2 &0.36$\pm$0.03& 9.03$\pm$0.06 & 18.3$\times$14.5 & 8.9$\times$5.2 & 1.6 & 5.0\\
238764 & 104& 7010.7$\pm$1.6 & 38.5$\pm$3.8 &0.25$\pm$0.02& 8.81$\pm$0.06  & 14.4$\times$11.3 & 5.9$\times$3.9 & 1.5 & 4.8 \\
248937 & 118& 8024.54$\pm$0.85 & 30.6$\pm$2.0 &0.60$\pm$0.04& 9.29$\pm$0.05 & 24.9$\times$18.5 & 13.2$\times$9.4 & 1.7 & 5.1 \\
248945 & 84& 5709.9$\pm$1.6 & 47.7$\pm$3.9 &0.37$\pm$0.03& 8.78$\pm$0.06 & 17.8$\times$14.4 & 6.4$\times$4.8 & 1.2 & 4.0 \\
334315 & 73&  5108.3$\pm$0.52& 46.4$\pm$1.3 &1.32$\pm$0.04& 9.22$\pm$0.05 & 39.1$\times$30.6 & 13.4$\times$10.0 & 1.0 & 7.3 \\
748738 & 56& 3896.5$\pm$0.9 & 29.2$\pm$2.1 &0.54$\pm$0.03& 8.61$\pm$0.06 & 24.7$\times$19.4 & 6.3$\times$4.7 & 0.80 & 4.4 \\
749251 & 106& 7262.2$\pm$0.8 & 33.6$\pm$1.9 &0.38$\pm$0.02& 9.01$\pm$0.05 & 17.9$\times$15.2 & 8.1$\times$6.5 & 1.5 & 4.1 \\ 
749290 & 97& 6512.5$\pm$1.0 & 42.7$\pm$2.5 &0.40$\pm$0.02& 8.95$\pm$0.05 & 19.8$\times$15.4 & 8.3$\times$6.1 & 1.4 & 5.2 
\enddata
\tablenotetext{a}{Distances calculated from ALFALFA flow model (see \citealp{haynes11a}).}
\tablenotetext{b}{\hi\ recessional velocity measured at 50\% of the peak flux; errors are statistical errors from the fit.}
\tablenotetext{c}{\hi\ line width measured at 50\% of the peak flux;  errors are statistical errors from the fit.}
\tablenotetext{d}{Semi-major and semi-minor axis of the \hi\  measured at a column density of $1.25\times10^{20}$atoms~cm$^{-2}$ as discussed in the text. $D_{\rm HI}$, is $2\times a$.}
\tablenotetext{e}{Semi-major and semi-minor axis as in note d, converted to physical units using the assumed distance.}
\tablenotetext{f}{Uncertainty in the semi-major axis, in physical units. Uncertainties for the semi-minor axis are similar.}
\tablenotetext{g}{Peak column density measured in the moment 0 images.}
\end{deluxetable*}

\section{Results}
\label{sec:results}


\subsection{The resolved \hi\ in HUDs}
\label{sec:results.hi}

\begin{figure*}[t!]
\centering
\begin{center}
{\bf \large AGC 248945} 
\vspace{-0.8em}
\end{center}
\includegraphics[width=0.30\textwidth]{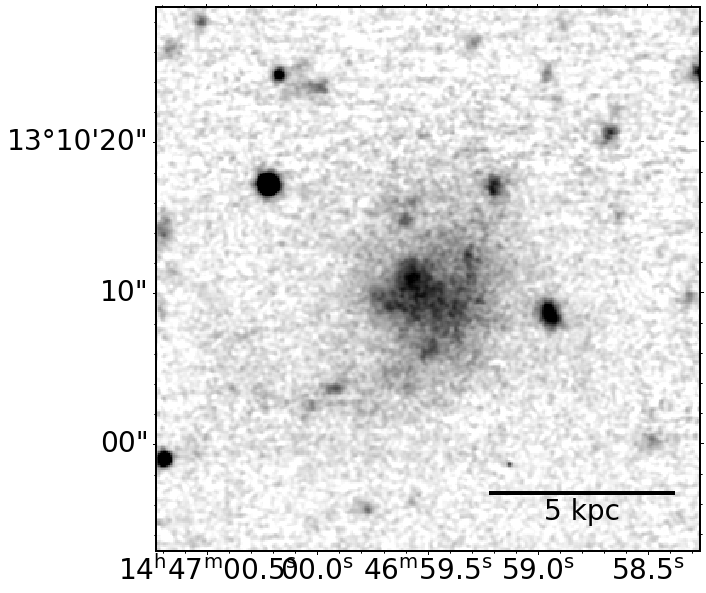}
\includegraphics[width=0.31\textwidth]{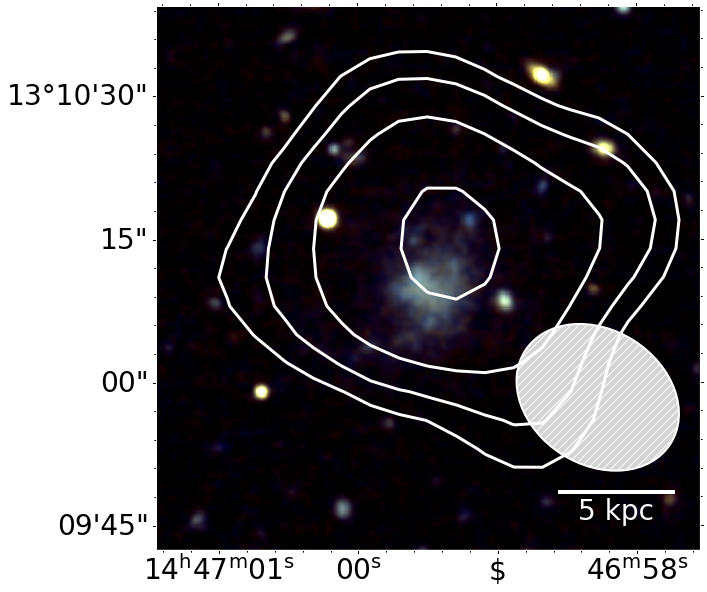}
\includegraphics[width=0.37\textwidth]{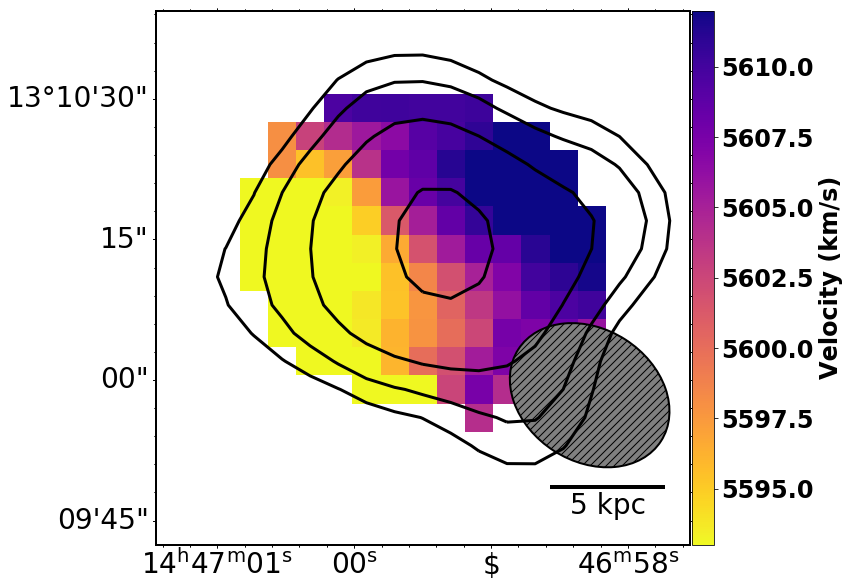}
\begin{center}
{\bf  \large AGC 749251}
\vspace{-0.8em}
\end{center}
\includegraphics[width=0.30\textwidth]{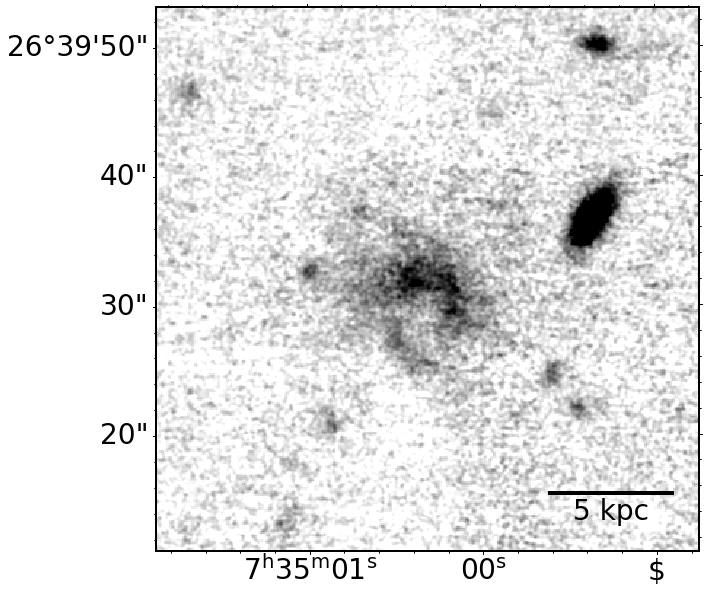}
\includegraphics[width=0.31\textwidth]{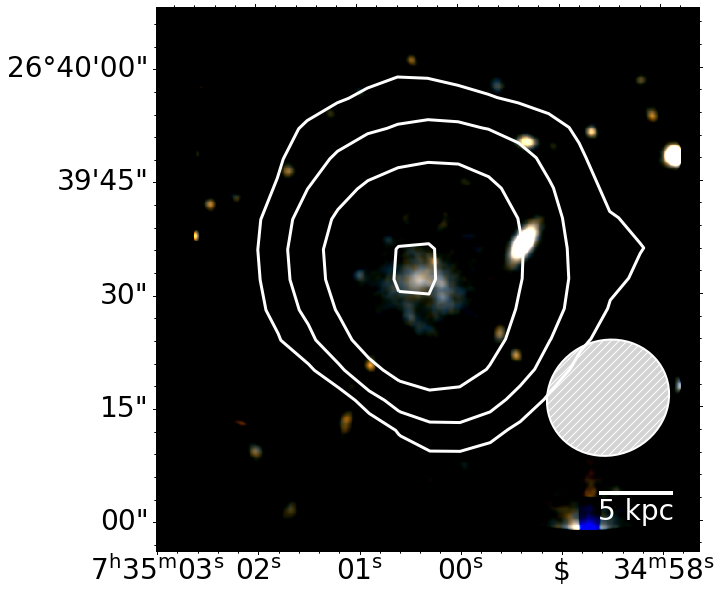}
\includegraphics[width=0.36\textwidth]{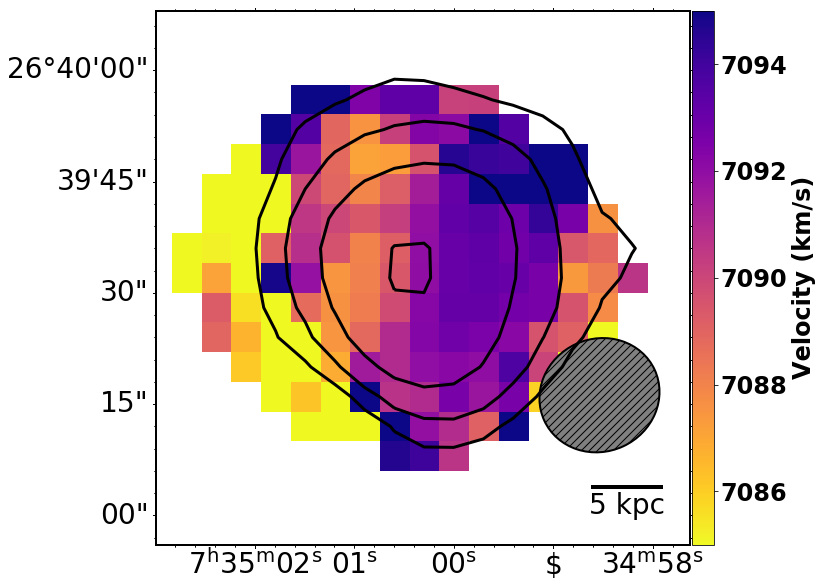}
\begin{center}
{\bf  \large AGC 748738}
\vspace{-0.8em}
\end{center}
\includegraphics[width=0.30\textwidth]{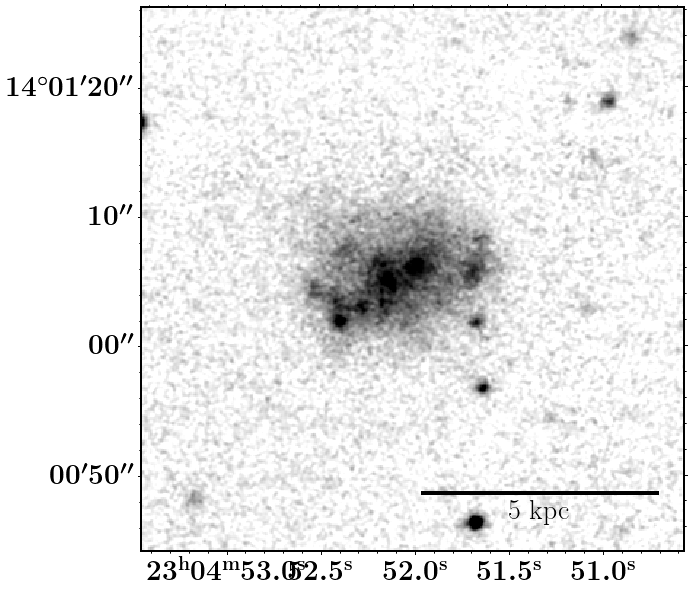}
\includegraphics[width=0.31\textwidth]{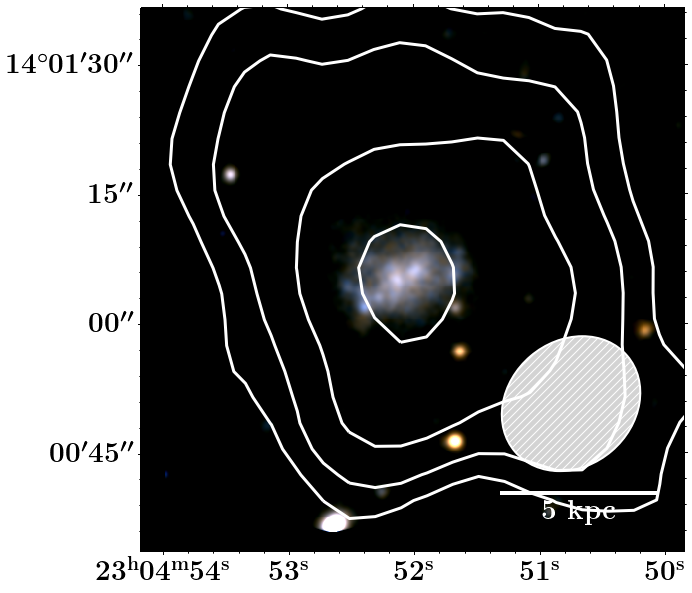}
\includegraphics[width=0.36\textwidth]{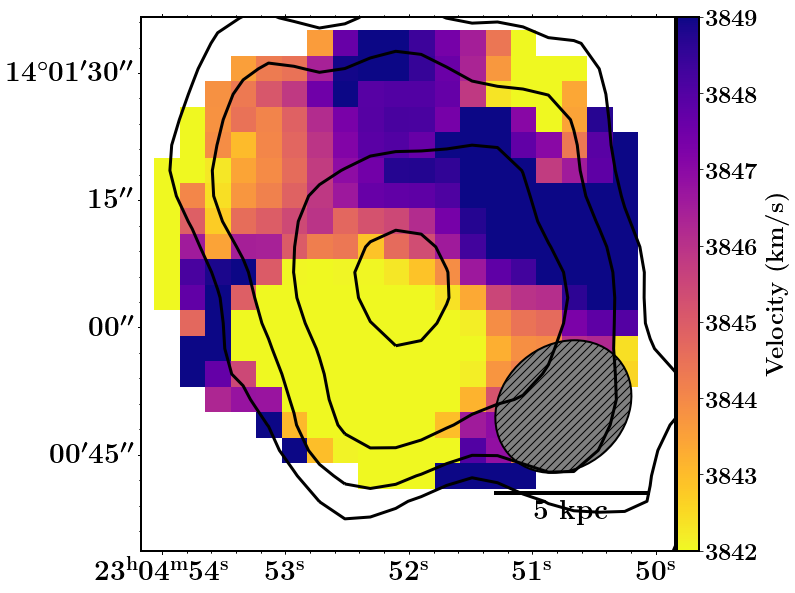} 
\vspace{-0.1cm}
\caption{Optical and velocity map comparisons for galaxies AGC 248945, AGC 749251, and AGC 748738 from top to bottom, showing clear velocity gradients across most galaxies and \hi\ gas extending well past the visible optical component. The far left column shows a zoomed in gray scale $g$-band optical image to show detail. The center column shows color optical images with \hi\ contours overlaid in white, and the right hand column shows \hi\ velocity maps, with column density contours shown in black. The contour overlays represented with the white and black lines in the center and right panels are at column density levels of 
0.43, 0.85, 1.7, and 3.4~$\times~10^{20}$ $\text{atoms~cm}^{-2}$
for each galaxy. The grey ellipses represent the size of the beam. The R.A. and decl. are in J2000 coordinates.
\label{fig:mmaps}
}
\end{figure*}

\begin{figure*}[t!]
\centering
\begin{center}
{\bf \large AGC 238764}
\vspace{-0.8em}
\end{center}
\includegraphics[width=0.30\textwidth]{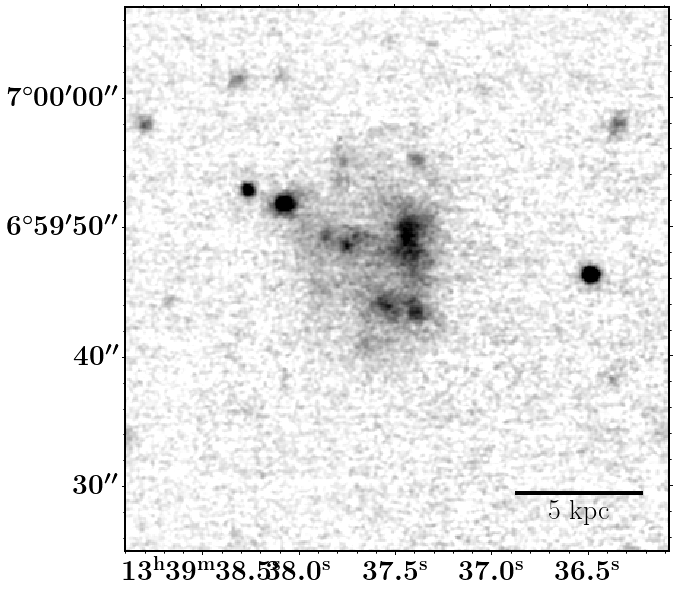}
\includegraphics[width=0.31\textwidth]{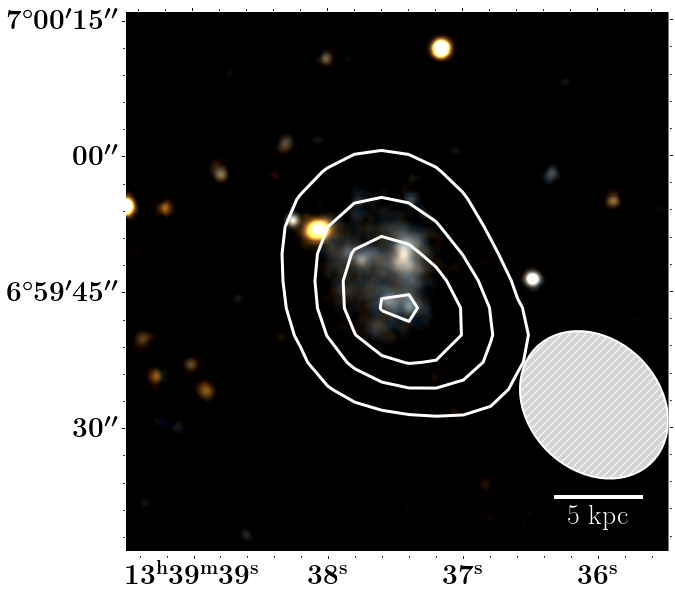}
\includegraphics[width=0.36\textwidth]{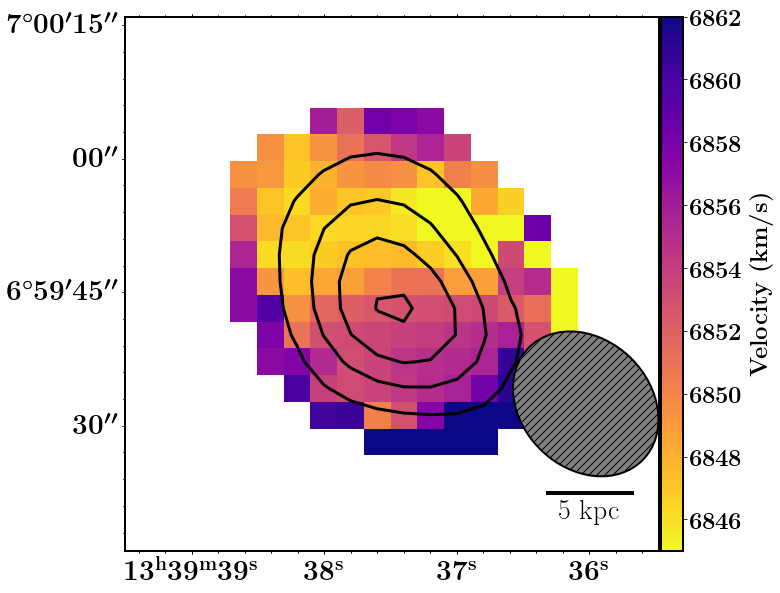}
\begin{center}
{\bf \large AGC 229110}
\vspace{-0.8em}
\end{center}
\includegraphics[width=0.30\textwidth]{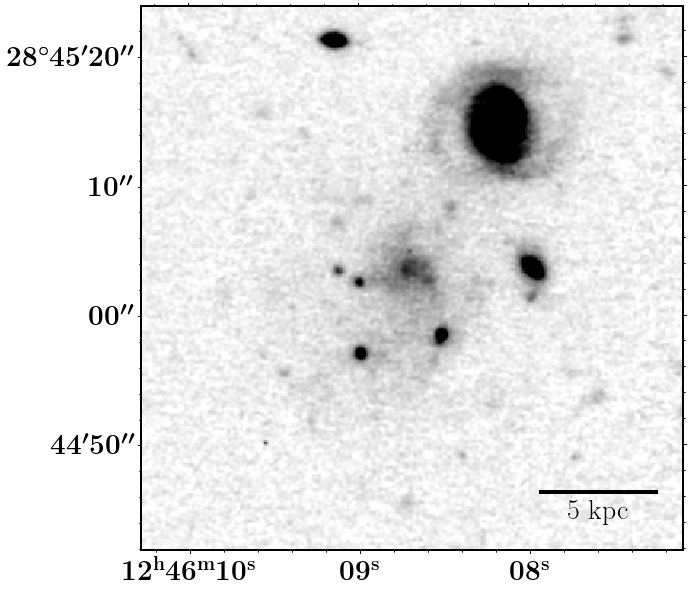}
\includegraphics[width=0.31\textwidth]{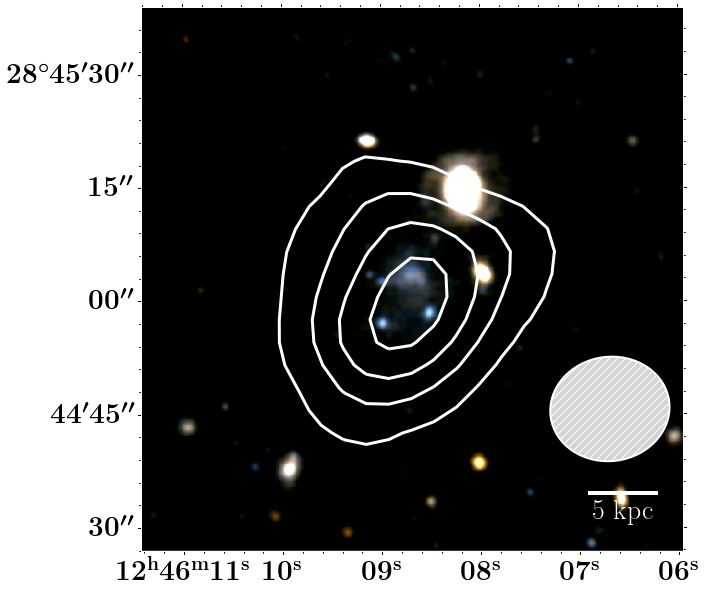}
\includegraphics[width=0.36\textwidth]{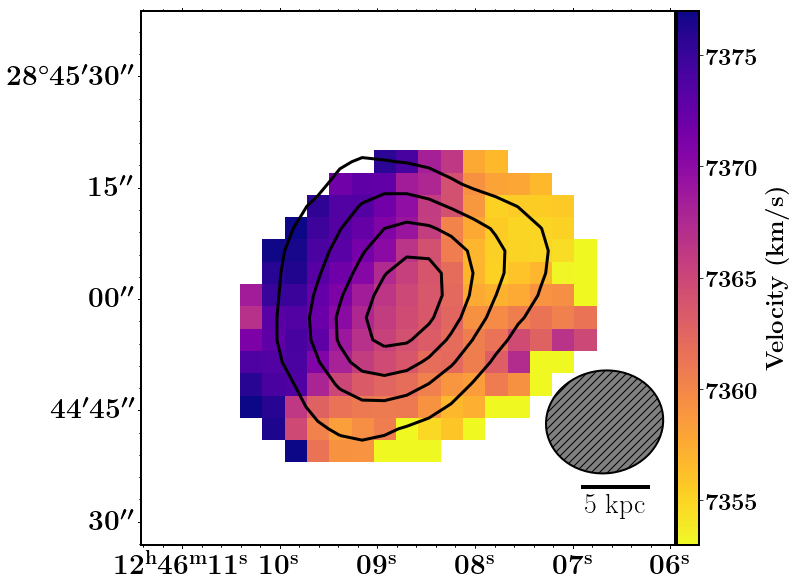}
\begin{center}
{\bf \large AGC 248937}
\vspace{-0.8em}
\end{center}
\includegraphics[width=0.30\textwidth]{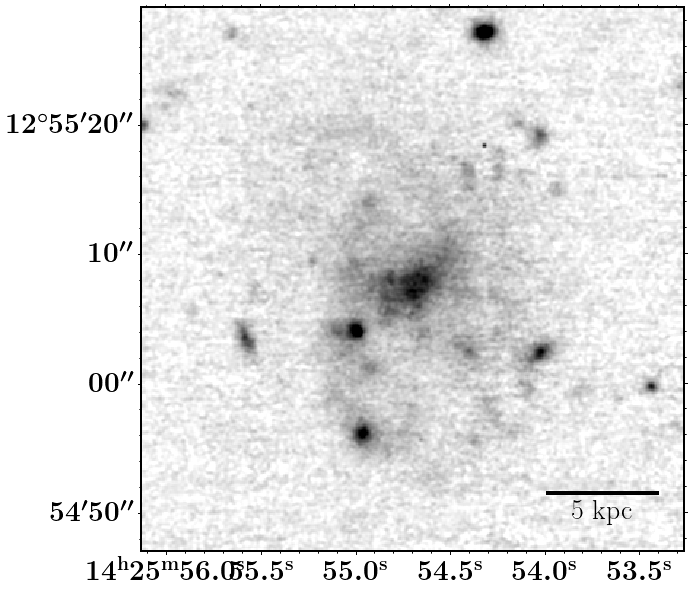}
\includegraphics[width=0.31\textwidth]{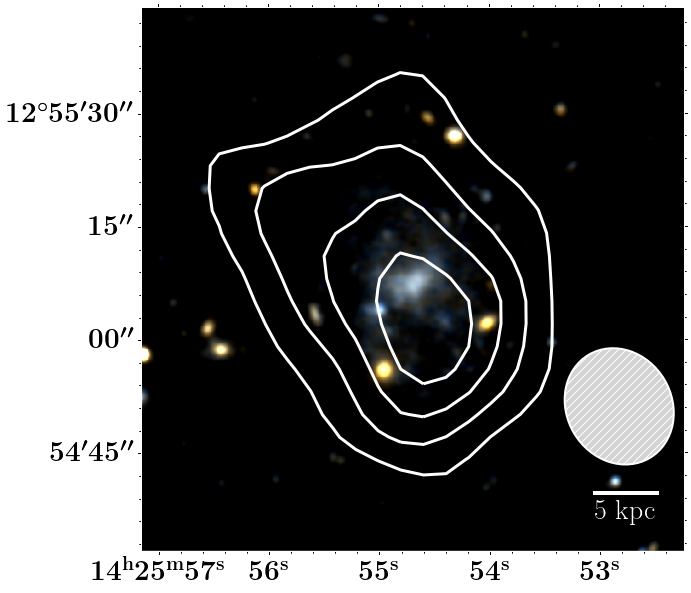}
\includegraphics[width=0.36\textwidth]{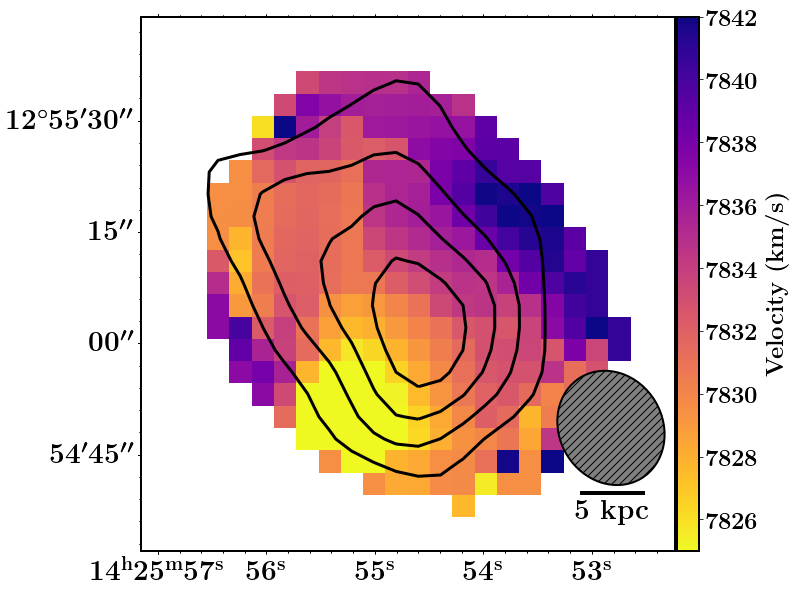}
\vspace{-0.1cm}
\caption{Optical and velocity map comparisons for galaxies AGC 238764, AGC 229110, and AGC 248937. The contour overlays represented with the white and black lines are at column density levels of 1.0, 2.0, 3.0, and 4.0~$\times~10^{20}$ $\text{atoms~cm}^{-2}$ for each galaxy. See Figure \ref{fig:mmaps} for details. 
\label{fig:mmaps1}
}
\end{figure*}

\begin{figure*}[t!]
\centering
\begin{center}
{\bf \large AGC 749290}
\vspace{-0.8em}
\end{center}
\includegraphics[width=0.30\textwidth]{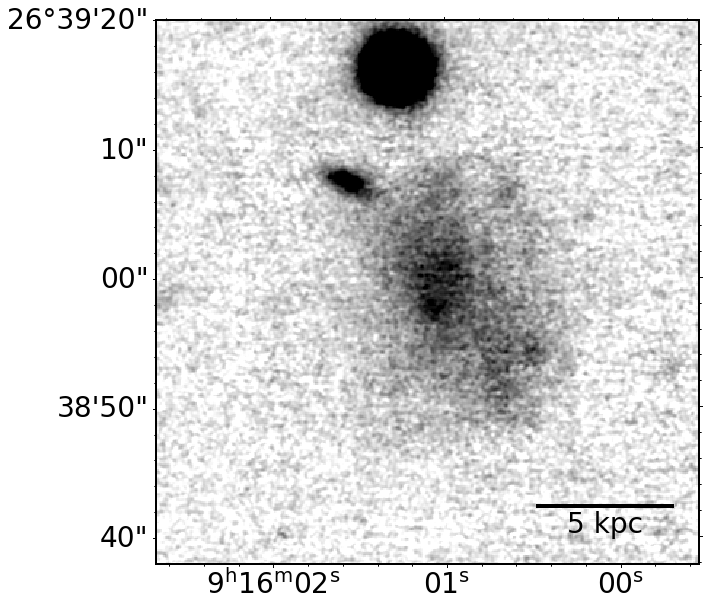}
\includegraphics[width=0.31\textwidth]{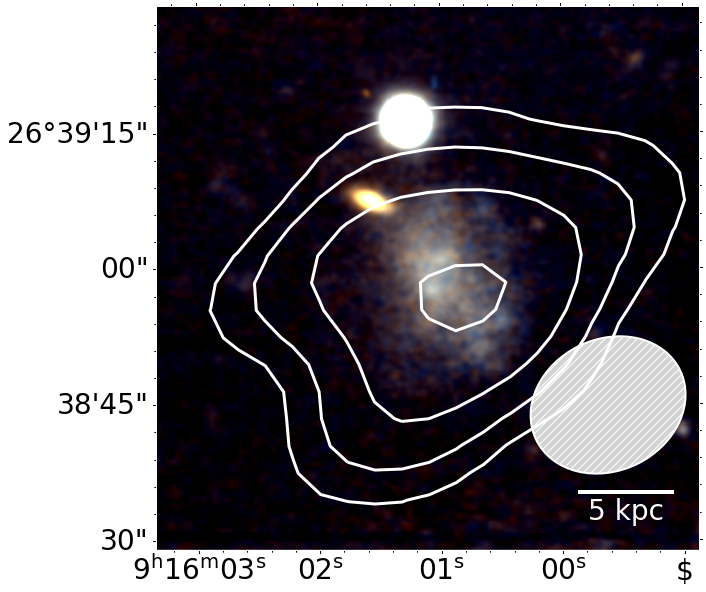}
\includegraphics[width=0.36\textwidth]{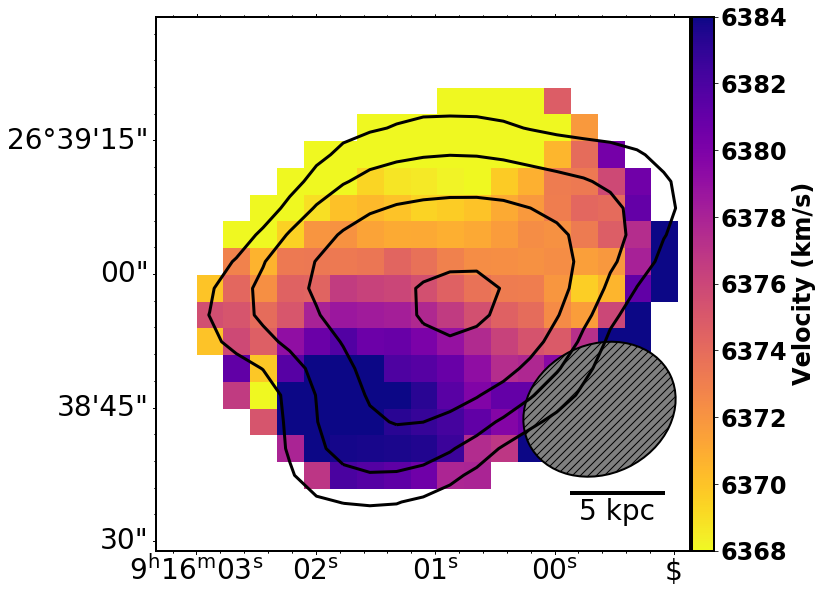}
\begin{center}
{\bf \large AGC 122966}
\vspace{-0.8em}
\end{center}
\includegraphics[width=0.30\textwidth]{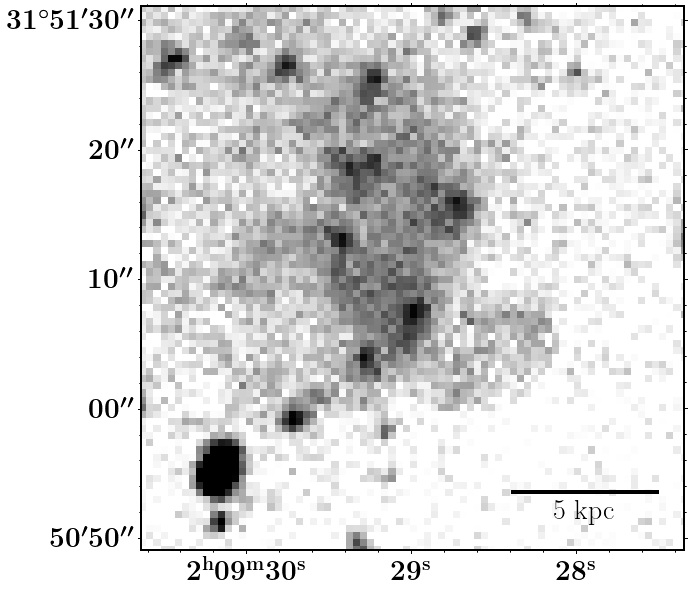}
\includegraphics[width=0.31\textwidth]{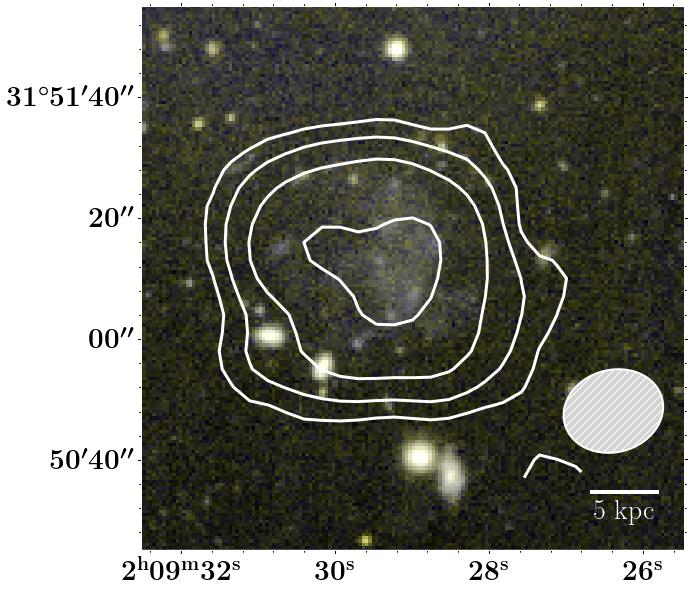}
\includegraphics[width=0.36\textwidth]{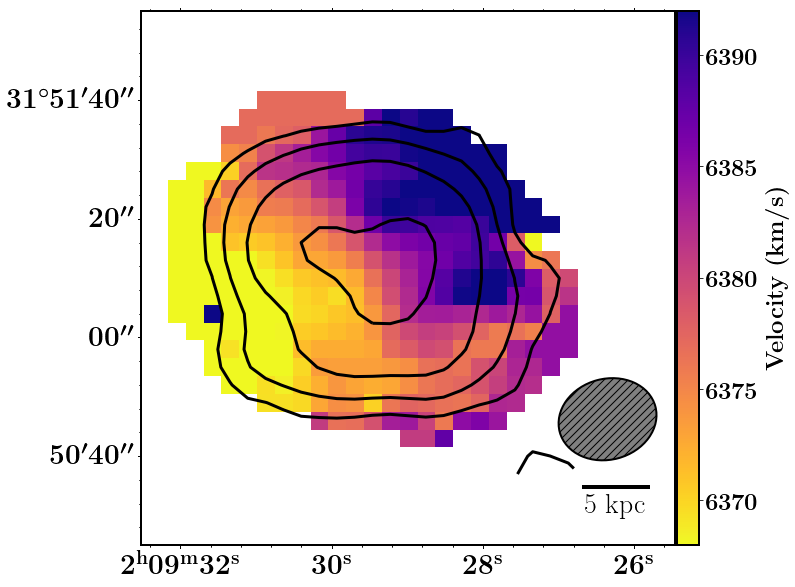}
\begin{center}
{\bf \large AGC 114905}
\vspace{-0.8em}
\end{center}
\includegraphics[width=0.30\textwidth]{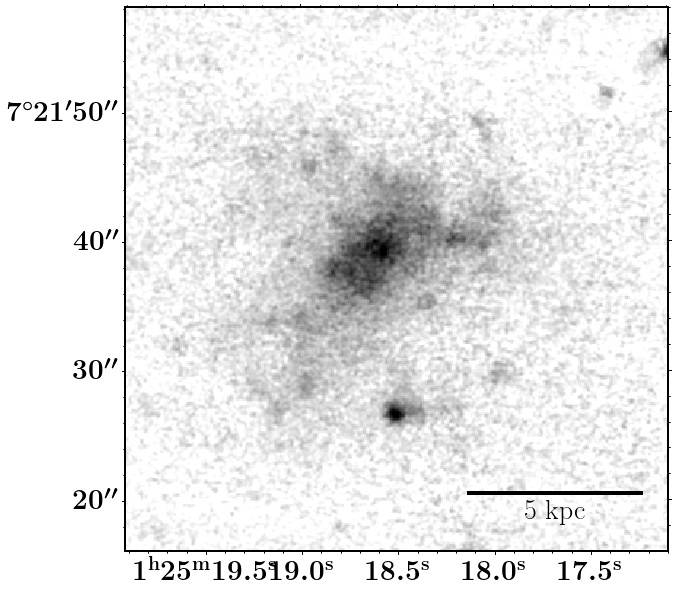}
\includegraphics[width=0.31\textwidth]{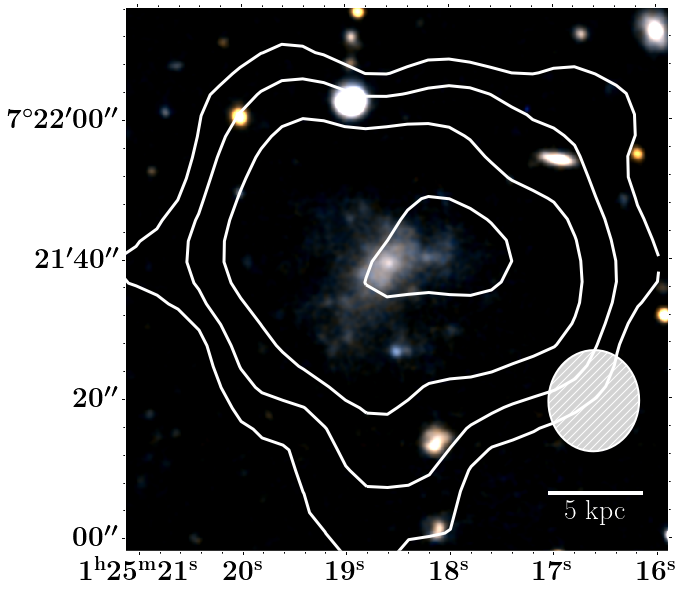}
\includegraphics[width=0.36\textwidth]{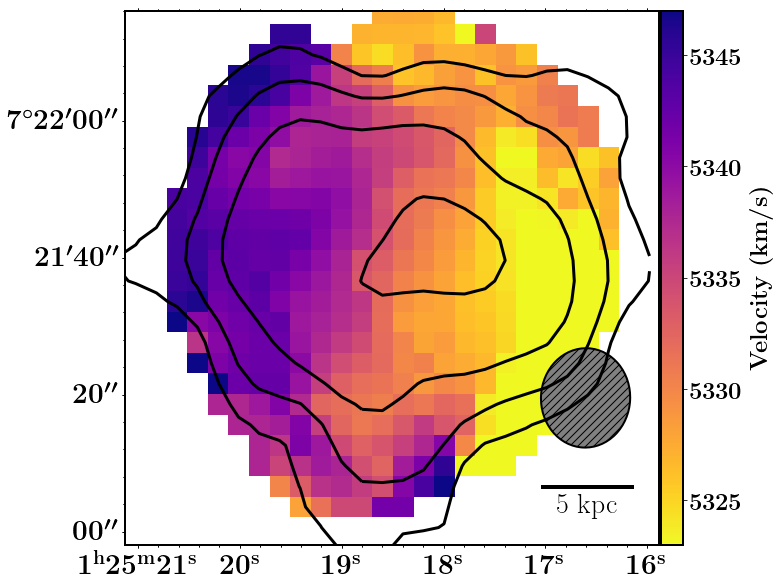}
\vspace{-0.1cm}
\caption{Optical and velocity map comparisons for galaxies AGC 749290, AGC 122966, and AGC 114905. The contour overlays represented with the white and black lines are at column density levels of 
0.56, 1.13, 2.25, and 4.50 $\times~10^{20}$ $\text{atoms~cm}^{-2}$ 
for each galaxy. See Figure \ref{fig:mmaps} for details. 
\label{fig:mmaps2}
}
\end{figure*}

\begin{figure*}[t!]
\centering
\begin{center}
{\bf \large AGC 219533}
\vspace{-0.8em}
\end{center}
\includegraphics[width=0.30\textwidth]{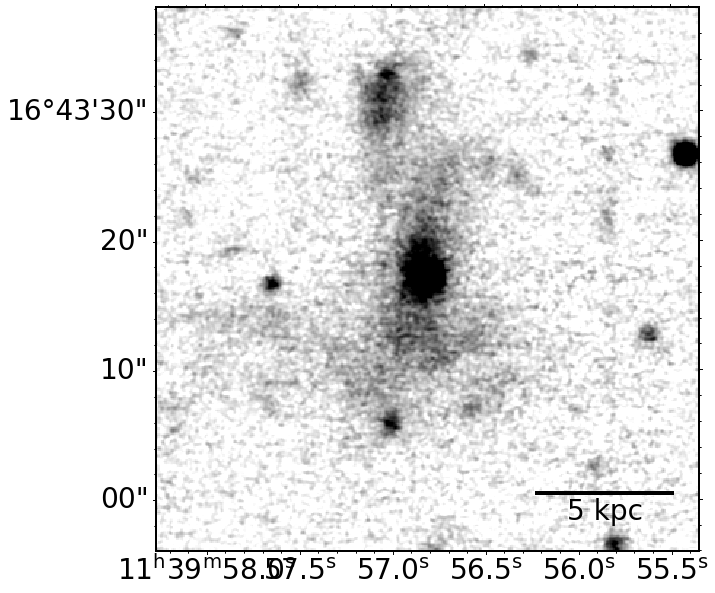}
\includegraphics[width=0.31\textwidth]{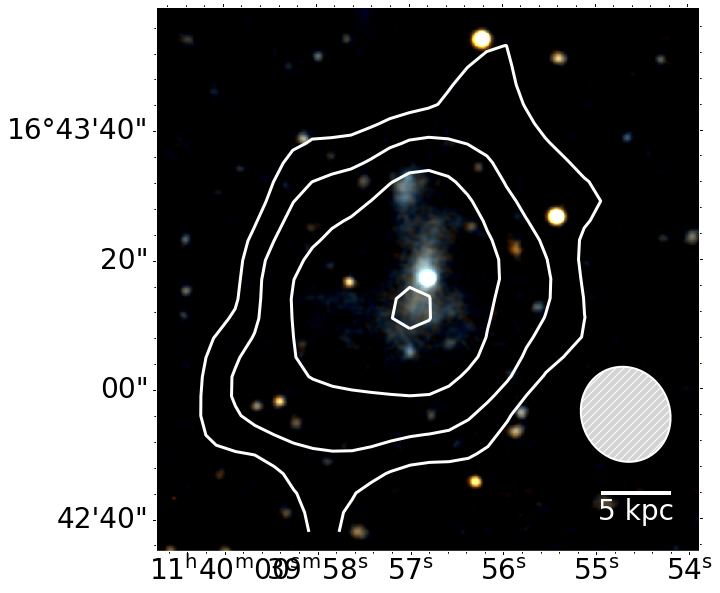}
\includegraphics[width=0.36\textwidth]{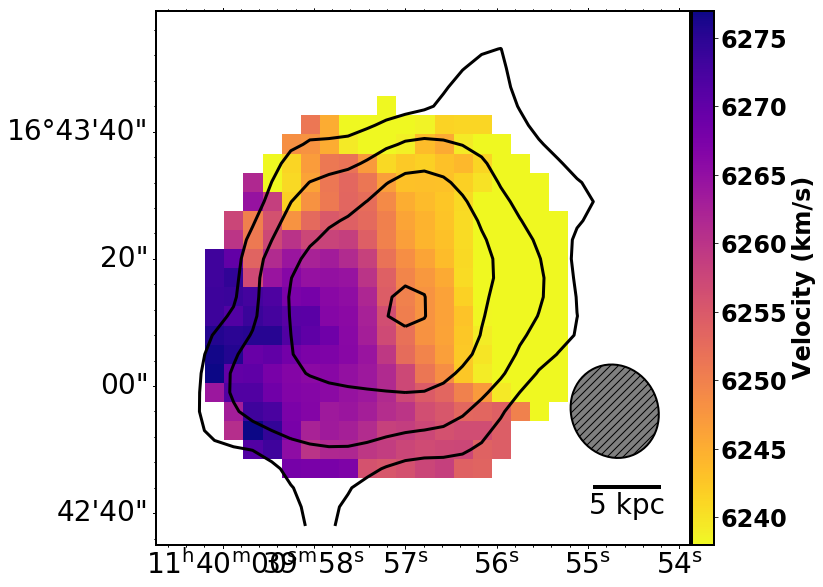}
\begin{center}
{\bf \large AGC 334315}
\vspace{-0.8em}
\end{center}
\includegraphics[width=0.30\textwidth]{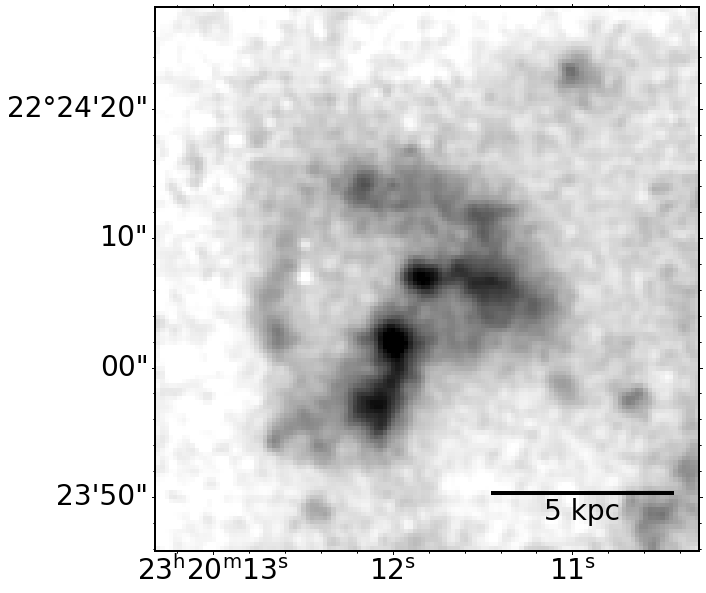}
\includegraphics[width=0.31\textwidth]{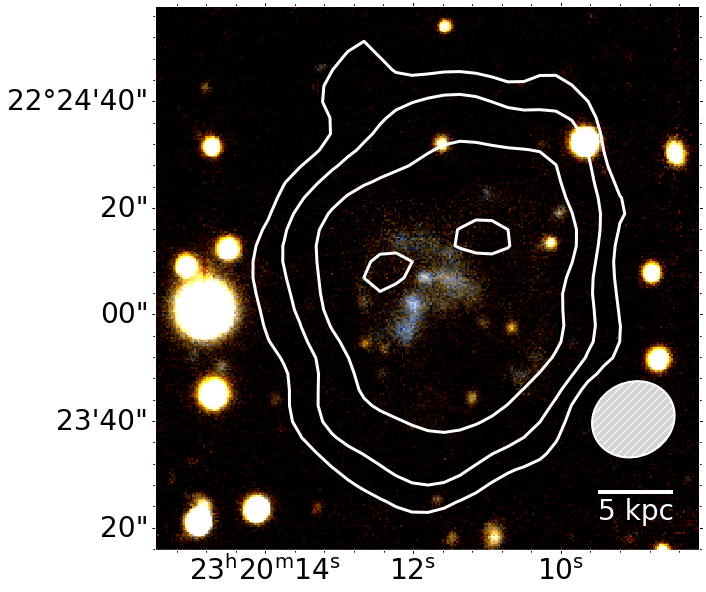}
\includegraphics[width=0.36\textwidth]{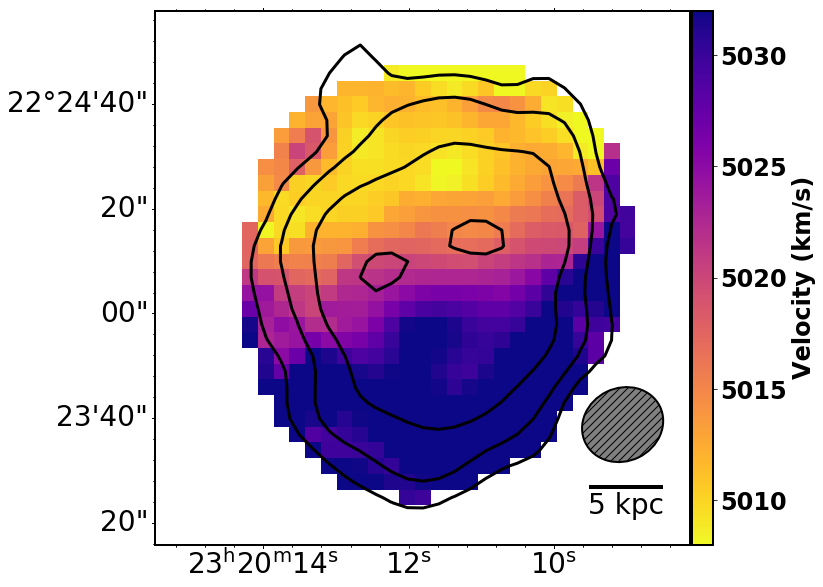}
\vspace{-0.1cm}
\caption{Optical and velocity map comparisons for galaxies AGC 219533 and AGC 334315. The contour overlays represented with the white and black lines are at column density levels of 
0.81, 1.63, 3.25, and 6.50~$\times~10^{20}$ $\text{atoms~cm}^{-2}$ 
for each galaxy. See Figure \ref{fig:mmaps} for details. 
\label{fig:mmaps3}
}
\end{figure*}

Figures \ref{fig:mmaps}-\ref{fig:mmaps3} show the data for the HUDs in our sample. The central panels show ODI color images of the optical components of the HUDs, with \hi\ column density contours overlaid as white lines, representing contours at the column density levels detailed in the figure captions. The figures are organized in order of ascending peak column density, with galaxies of similar peak column densities grouped with the same contour levels. The left-most column in each figure shows a zoomed-in grey-scale $g$-band ODI images of each galaxy. The right-hand panels show the moment 1 velocity maps, with the black lines representing the same contours as in the optical images. The grey circles represent the size of the beam for each source, and the color gradient represents the line of sight velocity in \kms. 

In the \hi\ maps, most sources are resolved with 3 or 4 independent beams across the full extent of detected \hi\ emission. 
Beam sizes range from 14.5\arcsec\ to 18.1\arcsec\ 
in major axis, which corresponds to physical resolutions from 4.6~kpc to 9.1~kpc at the distances of the sources (details listed in Table \ref{table:observations} and \ref{table:HIproperties}).

Sources AGC~238764 and AGC~248945 are only marginally resolved, with measured \hi\ semi-major axes at $1~\msun~\text{pc}^{-2}$ equal to or less than the beam size, 
and with only $\sim$2 beams across the full detected extent of the galaxies. These two sources are also the only two sources with recovered fluxes $F_{\rm VLA}/F_{\rm ALFA}$ well below 1 (0.55 and 0.59 respectively).  

Most HUDs in our sample appear to have regular \hi\ gas morphologies at our resolution. While this is to be expected at our relatively low physical resolution, it indicates these sources are not likely disturbed clouds of debris, and are presumably disks, allowing for reasonable estimates of the inclinations of these HUDs, which we discuss further in Section \ref{sec:discussion.inclin}.

All HUDs presented exhibit \hi\ gas distributions that extend well beyond their observed optical counterparts. This is characteristic of typical \hi-bearing galaxies (e.g., \citealp{boomsma08a,lelli16a}), but is interesting given the extended ``ultra-diffuse" nature of the stellar populations in question. We return to this point in Section \ref{sec:discussion.HImdr}. 
We also find that the gas is most dense in the center of these galaxies and becomes less dense as the \hi\ radius increases from the center. Peak \hi\ surface densities range from 3.98 to 7.50 x $10^{20}$ $\text{atoms~cm}^{-2}$, though it is important to note that this measurement assumes the gas fills the beam, and thus depends heavily on our resolution; the true peak density is almost certainly higher.

We estimate the \hi\ diameter at an \hi\ surface density of $1~\msun~{\rm pc}^{-2}$ along the major axis using a second moment analysis method, as described in \cite{banks95a}. 
Specifically, we measure the elliptical shape of emission
above the 1 $\msun~pc^{-2}$ level by applying a column density threshold of 1.25$\times10^{20}$~atoms~cm$^{-2}$ to the moment 0 map. We then calculate the first and second order moments of the 3\arcsec\ pixels at or above the threshold, which we then use to calculate the position angle and major and minor axes of the resulting fitted ellipse. Following \cite{wang16a}, we then correct the major axis diameter for beam smearing by:
\begin{equation}
    D_{\rm HI,0} = \sqrt{D_{\rm HI}^2-B_{\rm maj}\times B_{\rm min}}
\end{equation} 
and perform an analogous correction for the minor axis
to obtain the final values for the \hi\ sizes of our sources. The \hi\ radii measurements range from 6.7 kpc to 13.9 kpc along the major axis; details are reported in Table \ref{table:HIproperties} for each source.

The right-hand panels in Figures \ref{fig:mmaps}-\ref{fig:mmaps3} show the derived moment 1  maps, representing the velocity field for the galaxies in our sample, with the same total flux column density contours from the center panels overlaid in black to allow for direct comparison.\footnote{We note that the velocity maps are masked on smoothed images, which means that some of the noise outside the lowest signal-to-noise contour is displayed in each image.} We see velocity gradients and evidence of ordered rotation in the moment 1 maps for all of the sources. While there is little evidence for irregular motions in their \hi\ gas, some sources (e.g., AGC~248945 and AGC~334315) have very regular gradients, while others (e.g., AGC~749251 and AGC~748738) show gradients that are somewhat less regular, though the low resolution of our images does not allow for a more detailed analysis. 
In most cases, the orientation of the velocity gradient appears aligned along the \hi\ major axis, though, in three cases (AGC~229110, AGC~749251, AGC~749290), there is a moderate offset in measured position angles based on the kinematics and the morphology (see Section \ref{sec:discussion.inclin}).

Gaussian fitting to the global \hi\ line profiles extracted from the data cubes gives measured line widths for the HUDs that range from 29.2~\kms\ to 47.6~\kms, as listed in Table \ref{table:HIproperties}. As noted in \cite{leisman17a}, these velocity widths are lower than typically seen in galaxies of similar mass;  
\cite{mancerapina19b}, and \cite{mancerapina20a} show that when corrected for inclination (Section \ref{sec:discussion.inclin}), these sources fall off the baryonic Tully-Fisher relation (e.g., \citealp{mcgaugh00a,mcgaugh05a}). 


\subsection{The Stellar Populations of Resolved \hi-bearing UDGs}
\label{sec:results.optical}

The left-hand and center panels in Figures \ref{fig:mmaps}-\ref{fig:mmaps3} show the optical data for the 11 galaxies in our sample. The left-hand panels are $g$-band images with contrast chosen to highlight the low surface brightness features of the galaxies, while the center panels show color images with \hi\ contours at column density levels based on peak column density overlaid in white. 

All sources have very low surface brightness, with clearly detected stellar features that are nearly invisible in Sloan Digital Sky Survey imaging. Similar to the result reported in \cite{leisman17a}, most HUDs appear blue in color with mostly irregular morphologies without well defined features. Two sources, AGC~334315 and AGC~749251, have some arc-like features reminiscent of spiral arms, but without distinct patterns or strong evidence of a stellar disk. Three sources, AGC~114905, AGC~748738, and AGC~749290, have observed optical components that loosely resemble stellar disks in our optical images, but still would be better classified as irregular galaxies at the detected level. 
Also, one source, AGC~219533, has a bright, very blue higher surface brightness clump superimposed on its low surface brightness emission; it is not fully clear if it is a bright star forming region associated with the galaxy, or if it is a foreground or background source.

In many cases the centroid of the stellar population is aligned with the peak in the \hi\ column density, but there are several exceptions, including AGC~114905, AGC~238764, and AGC 248945. However, we note that AGC 238764 and AGC 248945 are less well resolved and lower signal-to-noise detections than the other sources in the sample, so the position of their peak column density is somewhat less well defined. 

Importantly, we note that we only poorly constrain the galaxies' inclinations using the stellar populations. 
Part of this is that the irregular morphologies make it difficult to accurately estimate the structural parameters used to constrain inclination and construct surface brightness profiles (section \ref{sec:data.optical}). While the process of azimuthally averaging the light tends to mitigate the effect of these structural uncertainties on the measurements of magnitude, surface brightness, and effective radius (especially as we hold the PA and axial ratio fixed), these uncertainties strongly affect the reliability of estimates of the major and minor axis. 
Moreover, the patchiness of the stellar light suggests that it may not be true that the apparent axial ratio of the light from the most visible stars actually traces the stellar disk, thus only providing at best a loose constraint on the systemic inclination.
 Only three of the sources have morphologies that loosely resemble stellar disks, AGC~114905, AGC~748738, and AGC~749290; for these three sources we find that attempts to measure their inclination optically always differs considerably from the value found using the \hi\ gas. Additionally, we note a significant difference between the optical and \hi\ position angles in AGC~749290. We come back to these points and their implications in Section \ref{sec:discussion.inclin}.  

\begin{figure*}[t!]
\centering
\includegraphics[width=0.9\textwidth]{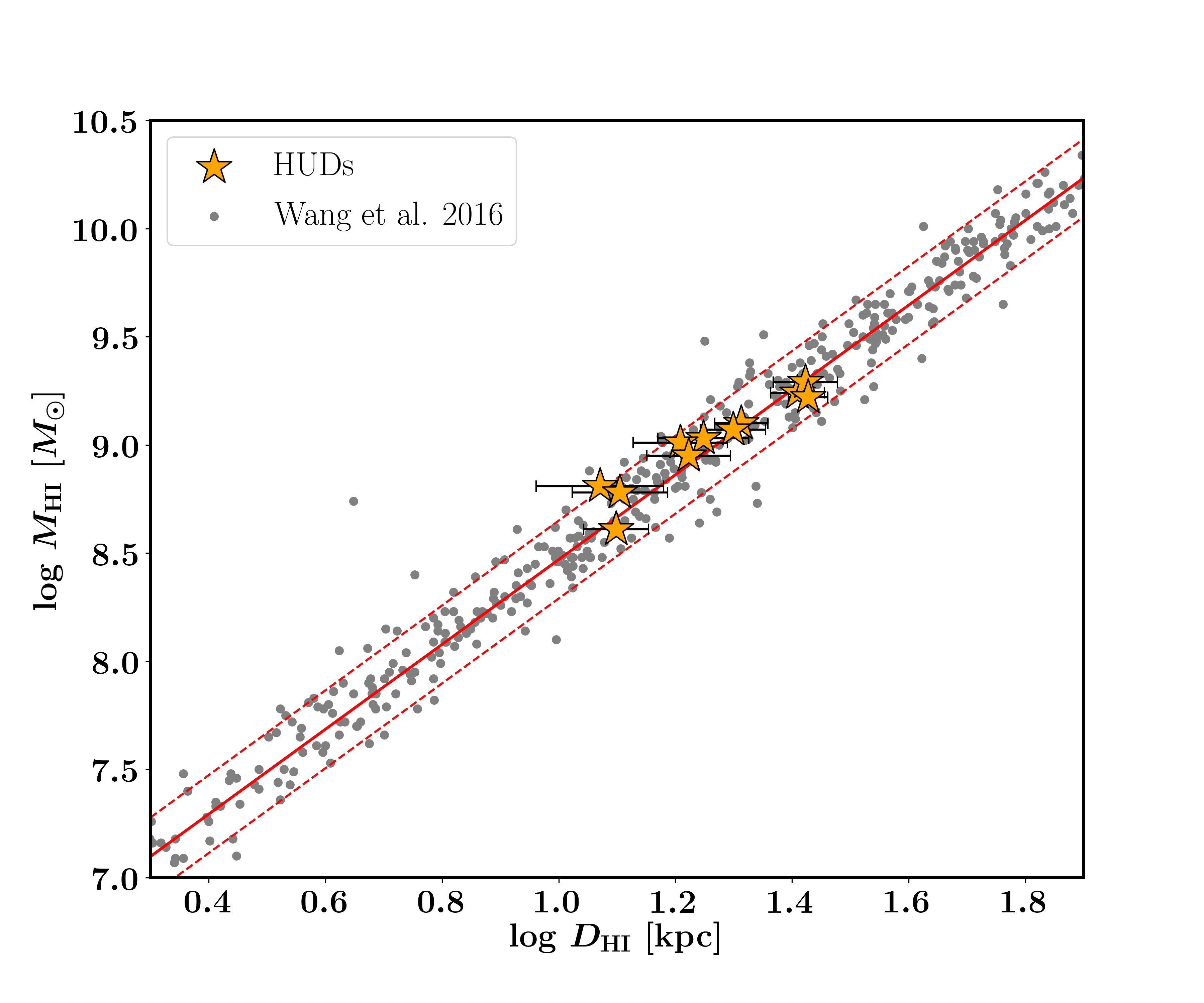}
\vspace{-0.1cm}
\caption{The $D_{\rm HI}-M_{\rm HI}$ relation, with the HUDs over plotted on data from \cite{wang16a}. The HUDs, despite having an extended, ``ultra-diffuse" stellar population, are \hi\ normal, with typical \hi\ radii and global gas densities. The solid line shows the linear fit from \cite{wang16a}. The dotted lines show the 3$\sigma$ scatter around the relation. \hi\ mass errors are smaller than the size of the markers.
\label{fig:HImd}
}
\end{figure*}

\subsection{\hi\ mass-diameter relation}
\label{sec:results:mhidhi}
The \hi\ mass-diameter relation is an observed tight correlation between the \hi\ mass of a galaxy
and its \hi\ radius measured at an 
isophotal level of $1~\msun~{\rm pc}^{-2}$
(e.g., \citealp{broeils97a,lelli16a,wang16a}).
Given that the HUDs are selected
to have extended optical counterparts, it is a natural
question to ask if they also have extended \hi\ disks for their mass.
Indeed, we may expect HUDs to fall off the relation, having larger \hi\ sizes for their mass if the formation mechanism responsible for their extended optical size also affects the gas disk (though the physical interpretation of the \hi\ mass-diameter relation is complex; see, e.g., \citealp{stevens19a}). If any of the HUDs are ``failed" L$_{\star}$ galaxies (e.g., \citealp{vandokkum15a,vandokkum16a}), an extended gas disk could help to explain their low surface brightness nature: 
an extended \hi\ disk for a given \hi\ mass implies a lower average surface density of gas, which could affect the ability of the neutral gas to condense and cool to form stars (e.g., \citealp{bacchini19a}). 
The \hi\ maps presented in Section \ref{sec:results} are sufficiently resolved to begin to address this
question, with the \hi\ disks typically resolved with 2-5 resolution elements across the whole disc. 

Figure \ref{fig:HImd} shows M$_{\rm HI}$-D$_{\rm HI}$ relation from \cite{wang16a}, with a red solid line indicating the linear fit to their relation: 
    $$\log D_{\rm HI} = (0.506 \pm 0.003) \log M_{\rm HI} - (3.293 \pm 0.009)$$
and with dashed lines showing their estimated 3$\sigma$ scatter ($\sigma \sim$0.06~dex) around the mean relation. Yellow stars show the HUDs from our sample plotted on the relation, with errors in the \hi\ diameters statistically estimated to be approximately 6\arcsec\ (the size of two pixels). These sources all fall on the relation; 
despite being selected for extended optical sizes, their \hi\ disks are indistinguishable from other galaxies of similar \hi\ mass.

We note that the \hi\ masses in this plot are measured from the VLA flux, which was consistent with the ALFALFA flux for all but two sources, as discussed in Section \ref{sec:results.hi}. If one instead uses the single dish fluxes for AGC~238764 and AGC~248945, they move up slightly ($\sim$0.2 dex), making them just barely consistent with the relation. 

We interpret this result further in Section \ref{sec:discussion.HImdr}.


\section{Discussion}
\label{sec:discussion}

Here we discuss our results, noting the importance of resolved \hi\ imaging for measuring inclinations, \hi\ radii, and column densities, as well as deep optical imaging for measuring optical sizes at low surface brightness. We explore the implications of the ultra-diffuse sources falling on the M$_{\rm HI}$-D$_{\rm HI}$ relation, 
finding that though their stellar disks are extremely low surface brightness, and though they are dark matter poor inside their disks \citep{mancerapina19b}, their \hi\ radii are typical for their optical effective radii. However, their \hi\ radii are large for their stellar mass, suggesting these galaxies are quite gas rich, consistent with the findings from \cite{leisman17a}. 

\subsection{\hi\ Inclination Measurements}
\label{sec:discussion.inclin}

\begin{deluxetable}{cccccc}
\tablecaption{\hi\ Morphological Parameters \label{table:inclinations}}
\tablecolumns{6}
\tablewidth{0pt}
\tablehead{
\colhead{AGC ID} &
\colhead{\hi\ PA\tablenotemark{a}} &
\colhead{\hi\ $i_{0.1}$\tablenotemark{b}} &
\colhead{\hi\ $i_{0.4}$\tablenotemark{c}} &
\colhead{\hi\ $i_{\rm MP}$\tablenotemark{d}} &
\colhead{Enclosed\tablenotemark{e}}\\[-0.8em]
\colhead{} &
\colhead{Degrees} &
\colhead{Degrees} &
\colhead{Degrees} &
\colhead{Degrees} &
\colhead{Fraction}
}
\startdata
114905 & 79 &	26$\pm$19 &	28$\pm$19 & 33$\pm$5 & 0.86$\pm$0.04 \\
122966\tablenotemark{*} & 67 &	39$\pm$15 & 43$\pm$15 & 34$\pm$5 & 0.87$\pm$0.05\\
219533 &  138 &	41$\pm$12 & 46$\pm$12 & 42$\pm$5 & 0.85$\pm$0.04\\
229110\tablenotemark{*} &  142 & 54$\pm$15 & 62$\pm$15 & -- & 0.77$\pm$0.07\\
238764 & 27 &	48$\pm$23 & 54$\pm$23 & -- & 0.76$\pm$0.08\\
248937\tablenotemark{*} & 22 &	44$\pm$13 & 49$\pm$13 &-- & 0.83$\pm$0.06\\
248945 & 268 &	41$\pm$20 & 46$\pm$20 & 66$\pm$5 &  0.78$\pm$0.07\\
334315 &  165 &	41$\pm$8 & 46$\pm$8 & 52$\pm$5 & 0.91$\pm$0.03\\
748738 & 30 &	41$\pm$14 & 45$\pm$14 &-- & 0.74$\pm$0.05\\
749251 &  5 &	36$\pm$23 & 40$\pm$23 &-- & 0.72$\pm$0.08\\
749290 &  118 &	44$\pm$17 & 49$\pm$17 & 39$\pm$5 &  0.79$\pm$0.06
\enddata
\tablenotetext{a}{Position angle estimated from the \hi\ morphological fitting. Uncertainties in the position angles are $\sim$6 degrees.}
\tablenotetext{b}{Morphological inclination estimated from the \hi\ axial ratio assuming q$_0$=0.1.}
\tablenotetext{c}{Morphological inclination assuming q$_0$=0.4}
\tablenotetext{d}{Inclination estimates from \cite{mancerapina19b}.} 
\tablenotetext{e}{The fraction of \hi\ flux within the measured \hi\ radius.}
\tablenotemark{*}{For these galaxies the observed  morphological axis does not coincide with the rotational gradient; the inclination estimates assuming nicely rotating disks should be treated with caution}
\end{deluxetable}

One of the main results from this sample, presented in detail in \cite{mancerapina19b} is that these isolated, very low surface brightness sources appear to be rotating too slowly for their baryonic mass, i.e., they lie off the baryonic Tully-Fisher relation. \cite{leisman17a} found low velocity widths for the larger parent sample, but resolved measurements are necessary to constrain the source inclinations and ensure against inclination selection bias. 

A number of methods can be employed for measuring the inclination of galactic disks (see, e.g., \citealp{garciagomez91a, andersen13a}).
\cite{mancerapina19b} determine inclinations by minimizing the residuals between observed moment 0 maps and model galaxies projected across a range of inclinations, including  a convolution step that makes the measurement of the inclination largely unbiased by the shape of the beam 
 (see \cite{mancerapina20a} for a detailed description). 

Another common method is to use ellipses fit to 2D \hi\ or optical photometric images to derive position angles and axial ratios, which are then converted to inclination assuming disks of some estimated thickness (see e.g., \citealp{boroson81a}). In the cases where the minor axis of our galaxies is resolved with at least two beams, we estimate axial ratios with our beam corrected major and minor axes measurements, and use these to compute inclinations using the standard formula (e.g., \citealp{jacoby92a}):
$$cos^2(i) = \frac{(b/a)^2-q_0^2}{1-q_0^2}$$
assuming two representative values for thin and thick disks, $q_0$=0.1 and 0.4. We report these, along with the measurements from \cite{mancerapina19b}, in Table \ref{table:inclinations}. 
For the galaxies included in both samples, our measurements are consistent with those derived from \cite{mancerapina19b} except in two cases: AGC~248945 and AGC~334315. These two galaxies had inclinations that were more than 10 degrees different (smaller) from those found in \citet{mancerapina19b}, though are still within 2$\sigma$ given uncertainties due to beam smearing.

We find that the estimated inclinations for our 11 sources tend to cluster around 40 degrees, ranging only from 26 to 54 degrees. Though our inclinations are only somewhat weakly constrained due to our comparatively low resolution, these constraints still give physically interesting results.
First, this inclination distribution seems inconsistent with the assumption of randomly oriented oblate spheroids. More specifically, we would expect randomly oriented oblate spheroids to be evenly distributed in bins of $\cos i$ (that is, for the number of galaxies in each inclination bin to increase from face on ($0^{\circ}$) to edge on ($90^{\circ}$) as a function of $\sin i$, see, e.g., \cite{binney98a}). This distribution can be modified somewhat by diameter and magnitude selection effects (e.g., \citealp{jones96a}), though these effects are modest for optically thin disks. Even with just 11 sources, our measured inclinations seem inconsistent with a random distribution (even accounting for magnitude or diameter selection effects); a KS test gives a probability that the sources in our sample have a random underlying distribution of $p<0.001$. 

That our galaxies are not randomly distributed is in itself, not surprising: the surface brightness criteria used in \cite{leisman17a}, along with the visual elimination of sources could easily bias the sample against edge on galaxies. This 
 may in part explain our observed distribution. 

However, while these results may suggest an inclination dependent selection bias in \cite{leisman17a} and \cite{janowiecki19a}, they also support the conclusion that their observed low velocity width distribution for the full sample of HUDs cannot be explained by inclination effects. A KS test comparing our observed inclination distribution with the distribution necessary to make the \cite{leisman17a} HUDs lie on the BTFR gives a probability $p=9\times10^{-7}$ of being drawn from the same distribution. This result thus supports the argument from \cite{mancerapina19b} that HUDs indeed do have abnormally low rotation velocities for their mass.

Yet, it is also possible that our observed inclination distribution indicates that the assumption of thin oblate spheroids may not be appropriate for this sample of galaxies. A number of authors (e.g., \citealp{vandenbergh88a}) have pointed out that dwarf irregular galaxies may not be well represented by spheroids with some intrinsic flattening, and may rather be triaxial, or at least have thicker disks. While the free fall time for \hi\ makes a thick \hi\ disk unlikely, a triaxial or otherwise irregular disk may be reasonable in light of their irregular stellar properties. Thus, we emphasize the importance of caution when approaching inclination measurements for these extreme sources.




As a further note of caution, we compare the \hi\ inclination estimates to estimates of optical inclination. As discussed in section \ref{sec:results.optical}, most sources in our sample have irregular optical morphologies, precluding accurate optical measurements of inclination.
However, for the three sources with the most regular morphologies in our sample, AGC 114905, AGC 748738, and AGC 749290, we estimate stellar inclinations to compare to our \hi\ measurements. In all three cases, the inclination estimate from the stellar population is $\sim20^{\circ}$ different from the estimates from \hi, with the stellar disk giving larger inclinations than the \hi\ (which, we note, would place them even further off the baryonic Tully-Fisher relation). We find the offsets in the HUDs sample are even more extreme than those in \cite{verheijen01a}, who also found a significant difference between inclinations measured optically, and using \hi. 

Further, there appears to be a $90^{\circ}$ difference between the position angle measurements for AGC 749290 with the optical being $289^{\circ}$ and \hi\ being $28^{\circ}$. 
These differences may be because the observed optical morphology in HUDs is likely dominated by patchy star formation.
Therefore, we reiterate that the stellar population of \hi\ -bearing UDGs is, in general, a poor indicator of inclination; we must also consider the \hi\ component when attempting to understand the orientation of HUDs. 

This note is especially important given the draw to inclination selection of edge-on HUDs (e.g., \citealp{he19a}), since this, in principle, removes the problem of inclination-correction for rotational (and surface brightness) studies. 
Since HUDs are likely a heterogeneous population including both irregular and spiral-like disks, it may or may not be that sources with high axial ratios are edge-on disks where inclination corrections can be ignored.
Our work confirms that the optical morphologies are not necessarily aligned with the kinematics of the gas, so one should have caution in interpreting unresolved kinematic information. 

\subsection{Implications of HUDs on the \hi\ Mass-Diameter Relation}
\label{sec:discussion.HImdr}

The results presented in Section \ref{sec:results:mhidhi} and Figure \ref{fig:HImd} show that the HUDs in this sample are all consistent with the \hi\ mass -- diameter relation of \citet{wang16a}. 
Another way of exploring this result is to calculate the average \hi\ surface density for our sample, which should give a result 
similar to other \hi-bearing galaxies, regardless of mass. Following  \cite{broeils97a}, 
we compute a characteristic surface density $\Sigma_{\rm HI,c} = 4\frac{M_{\rm HI}}{\pi {D_{\rm HI}}^2} = 4.15\pm0.27~\msun~pc^{-2}$, which is slightly higher than, but consistent with, their value of 3.8$\pm$0.1~$\msun~\text{pc}^{-2}$.\footnote{\citet{broeils97a} report a standard deviation of 1.1, which, for their sample of 108 galaxies gives a standard error on the mean of 0.11. The error on our characteristic surface density of 0.27 is the standard error on the mean.} 
Like other authors, we note that this characteristic $\Sigma_{\rm HI,c}$ is actually slightly higher than the actual average $\Sigma_{\rm HI}$, because we use the full measured \mhi\ rather than \mhi\ enclosed within D$_{\rm \sc{HI}}$ to most accurately compare with previous works.

Given the lower stellar surface density of these sources, the ``normal" \hi\ surface density may be somewhat surprising (e.g., \citealp{deblok96a}). 
One potential explanation is that the \hi\ radii are isophotal radii, measured at one position along the surface density profile, whereas the effective radius measurement used to define the optical radius is related to the slope of the light profile. Thus, these sources could still have larger than typical effective \hi\ radii if they have shallower slopes.

We explore this by measuring the flux enclosed within the \hi\ radius, compared with the flux outside the radius. We compute the fraction of the total flux enclosed within the \hi\ radius, and report the values in Table \ref{table:inclinations}. We find that our sources contain an average of $\sim81\%$ of the total \hi\ flux within the measured \hi\ radius, with a sample standard deviation 6\% (though we note the individual uncertainties based on the rms in our flux measurements are typically closer to 10\%). This is consistent with the estimates from \cite{wang16a} of $\sim83\pm10$\%. 

We also confirm that this technique can differentiate between samples with different slopes by applying this method to nine ATLAS 3D \citep{serra12a} sources used in the \cite{wang16a} sample which have significantly shallower \hi\ profiles (see \cite{wang16a} Figure 2). We find that these galaxies contain an average of 62\% of the total \hi\ flux within the \hi\ radii reported in \cite{wang16a}, with a standard deviation of 14\% (though we note that this comparison is somewhat limited since a number of the ATLAS 3D sources are disrupted, making estimation of the \hi\ radius more difficult (see \citealp{serra12a,serra14a}).     

To determine whether our masking procedure introduces a systematic bias, we also compute this enclosed fraction on our unmasked cubes. We find this decreases our average measured gas fraction to $76\pm9\%$, though most of this change is driven by two sources with additional positive flux at large radii in the unmasked cubes, which changes their estimated gas fractions by $\sim$16\%. If we remove these sources the average value is $79\pm6$\%. Though it provides a less direct comparison, we also note that we recover, on average, 96\% of the ALFALFA flux, so if we instead use ALFALFA fluxes, this could further decrease the gas fraction by an average of $\sim$4\%, still consistent with, but lower than the value measured by \cite{wang16a}. 

Thus, we suggest that the consistency of our measured enclosed fractions with \cite{wang16a} implies that the slope of the \hi\ profiles is not significantly different from typical \hi-bearing sources. However,  higher resolution, well resolved \hi\ profiles will be necessary to confirm this result. 

We also explore measuring optical isophotal radii for this sample of galaxies, and present the radii measured at 25~mag~arcsec$^{-2}$ in the $r$ band in Table \ref{table:optproperties}. However, while for typical gas-bearing galaxies radii measured at 25~mag~arcsec$^{-2}$ contain most of the optical light, for our galaxies, with peak surface brightness near 24~mag~arcsec$^{-2}$, this estimate misses most of the light of the galaxy. 
In fact, while for typical galaxies $r_{25}$ is 3-5$\times$ larger than the effective radius (i.e., the radius containing half the light), 
for our sources we find that $r_{25}$ is approximately 2/3 of the effective radius ($\langle r^{r}_{25} /r_{\rm eff}\rangle = 0.67\pm0.10$). 
This means that for our sample, measurements of R$_{\rm HI}$/$r_{\rm opt,25}$ are significantly elevated compared with other samples. We find $\langle$R$_{\rm HI}$/$r_{\rm opt,25}\rangle= 5.1\pm1.6$, compared to, e.g., $1.7\pm0.5$ as measured by \cite{broeils97a} for spiral galaxies, or $1.9\pm1.0$ from \cite{lelli16a} for the SPARC sample (disk galaxies).

\begin{figure}[t!]
\centering
\includegraphics[width=1.0\columnwidth]{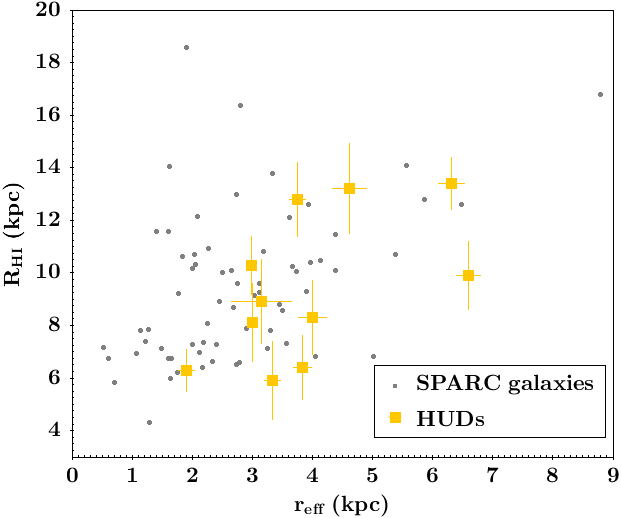}
\vspace{-0.1cm}
\caption{\hi\ radius (measured at 1 $\msun~pc^{-2}$) versus optical effective radius for our sample (yellow squares), compared with disk galaxies from the SPARC sample \citep{lelli16a}. The SPARC galaxies are selected to have the same \hi\ mass range, but have a wide range of stellar masses. The HUDs have elevated, but not extreme, stellar effective radii from an \hi\ perspective.
\label{fig:RHIvReff}
}
\end{figure}
 

If instead we compare \hi\ radii to optical effective radii, we find the effective radii are only somewhat larger than predicted by their \hi\ extent. Figure \ref{fig:RHIvReff} shows the \hi\ radii compared with the stellar effective radii for our sample and the SPARC sample of disk galaxies \citep{lelli16a}, selected to have the same \hi\ mass range (though a much larger stellar mass range). The HUDs fall in a similar parameter space, though with an elevated average effective radius (matching their initial selection). 
This is understandable, given that \cite{leisman17a} reported that these sources all have larger than usual gas fractions
(see, e.g., their Figure 4).

Thus, when considered from the \hi\ perspective, our sample of HUDs appears \hi\ normal at our current resolution, with slightly elevated stellar radii, but very small stellar masses and low surface brightnesses. 
HUDs do not significantly differ from typical galaxies in their average \hi\ surface densities; and at the resolution of our data, appear to have normal \hi\ gas distributions. 
While higher resolution observations will be required to confirm the full mass density profiles of these galaxies, the results presented here suggest that though their stellar populations are “ultra-diffuse,” the \hi\ gas in HUDs is not. 


%

\section{Conclusions}
\label{sec:conclusion}

In this paper we present deep optical and resolved \hi\ imaging of a sample of 11 isolated \hi-bearing ultra-diffuse galaxies selected from the ALFALFA survey. These sources are notable because of their extreme low surface brightnesses, large optical sizes, and their relative isolation from other sources. 

We find that the sources have irregular optical morphologies with generally blue colors, consistent with the findings in \cite{leisman17a}. We note that the stellar populations in these irregular sources, perhaps not unexpectedly, often give inconsistent inclination measurements with the \hi; thus resolved \hi\ measurements are essential for inclination-dependent studies. However we also caution that our observed inclination distribution seems inconsistent with a random distribution, potentially due to selection effects, or \hi\ distributions that deviate from circular disks.

We further find that at the level of our resolution, the sources in this sample appear to have normal \hi\ morphologies, with the \hi\ extending beyond the observed diffuse stellar population in all cases.

We plot these extreme sources on the M$_{\rm HI}$-D$_{\rm HI}$ relation, and find that they are all consistent with the relation, and that they have normal global \hi\ surface densities. We explore the gas fraction enclosed within the \hi\ radius, and find it consistent with normal \hi\ disks, possibly suggesting that though their stellar populations are ultra diffuse, the \hi\ in HUDs is not.
This suggests that globally, their extreme surface brightness may not be driven by 
low or anomalous \hi\ densities or distributions, though higher resolution observations will be necessary to confirm this suggestion. 

This publicly available,\footnote{Final reduced \hi\ and optical data products are available at \url{https://scholar.valpo.edu/phys_astro_fac_pub/186/}} rich data set provides an important baseline for future exploration of these mechanisms and other questions relying on resolved studies of HUDs (e.g., their dark matter content,  \citealp{mancerapina19b}), and for direct comparisons between theoretical models and observations.

\acknowledgments

{\bf Acknowledgments}.  The authors acknowledge the work of the entire ALFALFA
collaboration in observing, flagging, and extracting sources, and would like to thank the anonymous referee for useful suggestions that improved the quality of the paper. 
The ALFALFA team at Cornell has been supported by grants NSF/AST 0607007, 1107390 and 1714828 and by the Brinson Foundation.
PEMP and FF are supported by the Netherlands Research School for Astronomy (Nederlandse Onderzoekschool voor Astronomie, NOVA), Phase-5 research programme Network 1, Project 10.1.5.6.
EAKA is supported by the WISE research programme, which is financed by the Netherlands Organization for Scientific Research (NWO).
Worked carried out by NJS, HJP, and KLR for this project is supported by NSF grant AST-1615483. 
LG and LL acknowledge support from Valparaiso University.

JMC is supported by NSF/AST-2009894. JMC and QS acknowledge support from Macalester College.  

This work is based in part on observations made with the VLA, Arecibo Observatory, and WSRT.
The VLA is a facility of the National Radio Astronomy Observatory (NRAO). NRAO is a facility of the National Science Foundation operated under cooperative agreement by Associated Universities, Inc.  The Arecibo Observatory is operated by SRI International under a cooperative agreement with the National Science Foundation (AST-1100968), and in alliance with Ana G. M\`endez-Universidad Metropolitana, and the Universities Space Research Association. 
The Westerbork Synthesis Radio Telescope is operated by the ASTRON (Netherlands Institute for Radio Astronomy) with support from NWO.

Observation presented in this paper were obtained with the WIYN Observatory, which is a joint facility of the NSF’s National
Optical-Infrared Astronomy Research Laboratory, Indiana University,
the University of Wisconsin-Madison, Pennsylvania State University,
the University of Missouri, the University of California-Irvine and
Purdue University.
\bibliography{mybib}
\bibliographystyle{aasjournal}



\end{document}